\documentclass[12pt, onecolumn, draftcls]{IEEEtran}
\usepackage[marginal]{footmisc}
\usepackage{float}
\usepackage{graphicx}
\usepackage{subfig}
\usepackage{array}
\newcommand{\PreserveBackslash}[1]{\let\temp=\\#1\let\\=\temp}
\newcolumntype{C}[1]{>{\PreserveBackslash\centering}p{#1}}
\newcolumntype{R}[1]{>{\PreserveBackslash\raggedleft}p{#1}}
\newcolumntype{L}[1]{>{\PreserveBackslash\raggedright}p{#1}}
\usepackage{amssymb}
\usepackage{amsmath}
\usepackage{amsthm}
\usepackage{amsfonts}
\usepackage{mathrsfs}
\usepackage{algorithm}
\usepackage{algorithmic}
\usepackage[no adjust]{cite}
\usepackage{enumerate}

\usepackage{caption}
\usepackage{booktabs}
\usepackage{setspace}

\usepackage[T1]{fontenc}% optional T1 font encoding

\usepackage{amsmath,algorithm,algorithmic,array,amssymb,amsthm,amsfonts}

\usepackage{booktabs,bm,boxedminipage,cite,color,dsfont,epsfig,setspace,fancyhdr,float, graphicx, url}
\usepackage{multirow}

\usepackage[cmintegrals]{newtxmath}

\usepackage{amsmath,amsthm,amssymb}

\begin{document}

\title{\Large Multiuser Video Streaming Rate Adaptation: A Physical Layer Resource-Aware Deep Reinforcement Learning Approach}

\author{\small Kexin~Tang,~\IEEEmembership{\small Member,~IEEE}, \small Nuowen~Kan, \small Junni~Zou,~\IEEEmembership{\small Member,~IEEE}, Xiao~Fu, Mingyi~Hong, and Hongkai~Xiong,~\IEEEmembership{\small Senior Member,~IEEE}

\thanks{K. Tang, N. Kan, J. Zou and H. Xiong are with Shanghai Jiao Tong University, P.R. China. M. Hong is with University of Minnesota, USA. X. Fu is with Oregon State University, USA.}
}

\maketitle
%\vspace{-2.cm}
\pagestyle{empty} % no page number for the second and the later pages
\thispagestyle{empty} % no page number for the first page
\begin{abstract}
\renewcommand{\baselinestretch}{1.6}
We consider a multi-user video streaming service optimization problem over a time-varying and mutually interfering multi-cell wireless network. 
The key research challenge is to appropriately adapt each user's video streaming rate according to the radio frequency environment (e.g., channel fading and interference level) and service demands (e.g., play request), so that the users' long-term experience for watching videos can be optimized.

%(e.g., channel states, interference level, buffer status)

To address the above challenge, we propose a novel two-level cross-layer optimization framework for multiuser adaptive video streaming over wireless networks. The key idea is to jointly design the physical layer optimization-based beamforming scheme (performed at the base stations) and the application layer Deep Reinforcement Learning (DRL)-based scheme (performed at the user terminals), so that a highly complex multi-user, cross-layer, time-varying video streaming problem can be decomposed into relatively simple problems and solved effectively. Our strategy represents a significant departure for the existing schemes where either short-term user experience optimization is considered, or only single-user point-to-point long-term optimization is considered. Extensive simulations based on real-data sets show that the proposed cross-layer design is effective and promising.
\end{abstract}

\begin{IEEEkeywords}
%\vspace{-1.0cm}
Video streaming over wireless cellular networks, beamforming, rate adaptation, reinforcement learning, cross-layer design.
\end{IEEEkeywords}

\begin{center}
\small
\end{center}
\newpage
\IEEEpeerreviewmaketitle
\hyphenation{op-tical net-works semi-conduc-tor}

\section{Introduction}

\IEEEPARstart{M}{o}bile video streaming has become a major drive for the continued growth of wireless network data traffic: It has already accounted for 60\% of the total mobile data traffic in 2016, and this number is projected to be increased to 78\% by 2021 \cite{Cisio}. This trend imposes significant challenges to the task of real-time delivery of resource-demanding video streams over wireless networks.

%\noindent{\bf Adaptive Streaming.}
To ensure high quality of experience (QoE) for mobile users, while coping with  ever growing user heterogeneity, (in terms of demands for video content, display devices and available network resources),  and fast changing network conditions, flexible video content delivery techniques such as Dynamic Adaptive Streaming over HTTP (DASH) have been developed~\cite{DASH}. %In particular, DASH has been developed to adapt the video quality to the changing network conditions and/or user device capabilities. 
In DASH, each video is encoded at different bitrates and/or resolutions to generate several representations. These representations is further divided into small chunks containing a few seconds of video content (typically 2 seconds). This way, client-side video players are able to dynamically select and switch representations at each chunk boundary to fit users' network situations. A variety of rate adaptation techniques~\cite{rate-based, buffer-based, MPC} have been proposed to improve the user-perceived QoE  based on throughput and/or buffer status locally observed on the user side. 

However, these techniques are mostly designed for single-user settings. When {\it multiple} DASH  clients are competing for the network resources,  it has been shown that these algorithms may result in video bitrate oscillation, network resource underutilization, or QoE unfairness among users \cite{huang, shape, PANDA}.  Particularly, in mobile networks, a DASH user adjusts its video quality based on the measured throughput that actually depends on its transmission rate. Since the transmission rate for each user is determined by physical layer wireless resource allocation schemes, application-agnostic and/or unfair resource allocation in the bottom layer inevitably leads to inferior QoE in the application layer. 

The {\it overarching goal} of this paper is to design an effective cross-layer approach that is capable of meeting the following goals: 1) Achieving both {\it fair wireless resource allocation} in the physical layer, and  {\it adaptive bitrate selection} in the  application layer; 2) Obtaining effective and high-quality video content delivery among heterogeneous users, and over time-varying wireless networks.

\subsection{Related Literature}

\noindent{\bf Rate Adaptation Techniques.}
HTTP-based adaptive streaming (standardized as DASH~\cite{DASH}) is a client-driven video streaming protocol that allows the users to retrieve the video chunks with the desired bitrate by sending HTTP requests. Naturally, most of the existing rate adaptation algorithms\cite{rate-based,buffer-based,MPC} are carried out at the user side. However, these algorithms have inferior performance (e.g., lower video quality, frequent quality switches, QoE unfairness) in the presence of multiple competing video flows. To deal with these issues, various client-side~\cite{PANDA, FESTIVE} and server-side~\cite{shape, globecom} solutions have been proposed to develop robust DASH systems in the multiuser scenario. 

Jiang \textit{et al.}~\cite{FESTIVE} propose a client-side rate adaptation algorithm (FESTIVE) to achieve reasonable tradeoff between three conflicting objectives: efficiency, fairness and stability. The authors in~\cite{PANDA} design the PANDA (Probe and Adapt) algorithm to probe the fair share bandwidth based on the measured TCP throughput. However, these two client-side algorithms tend to be conservative for video bitrate selection, leading to bandwidth underutilization unless the network conditions are extremely steady. A server-based traffic shaping method is proposed in~\cite{shape} to reduce the video bitrate oscillations by limiting the throughput for each chunk to its encoding rate. Marai \textit{et al.}~\cite{globecom} propose a server-side rate modification mechanism (SO-DASH) that decreases the requested bitrates of users who dominate the bottleneck bandwidth. Since the server needs to maintain a throughput record for each user, server-based methods have poor scalability as the number of DASH users increases. We refer the reader to~\cite{survey} for a comprehensive review of rate adaptation techniques for DASH.

\noindent{\bf Cross-Layer Techniques.}  In order to ensure fairness among DASH clients competing for the network resource, some cross-layer resource allocation approaches for multiuser adaptive video streaming~\cite{crosslayer1, crosslayer2, crosslayer3, crosslayer4, crosslayer5} have been proposed to jointly optimize the physical layer transmission rate for each user with QoE/bandwidth fairness objectives, and then overwrite the requested video bitrate of users by a network-side centralized proxy server to match the optimization result. For instance, the video bitrate will be reduced for users whose requested bitrate is larger than the transmission rate allocated to it. 
Note that these methods are less appealing as centralized operations are usually undesired for upper layers due to privacy/security reasons and the asynchronous nature of user requests.
%Nonetheless, these methods are less appealing in practical applications due to the need of a centralized proxy server to modify the users' bitrates. %This violates the original intention of adaptive streaming systems, which regards the external network environment as a black box and distributively execute rate adaptation logic at each user side to adapt to network changes. 
More importantly, these cross-layer resource allocation algorithms merely focus on short-term QoE maximization problems (within a time slot). 
In practice, however, it is much more preferable to offer high and fair QoE to users over a long period of time, since the event of video watching can easily span hours.
In addition, such short-term optimization strategies normally lack the ``looking ahead of time'' (i.e., prediction) ability, which is important for maintaining high QoE in video streaming (e.g., increasing buffer occupancy for predicted link deterioration).

\noindent{\bf Reinforcement Learning-based Techniques.} To better capture the time-varying aspect of the video streaming application, and to account for the user's overall experience, it is usually desirable that some kind of {\it long-term} utility function is optimized for video streaming applications.  As a general framework for dynamic control (decision) problem, reinforcement learning (RL)~\cite{RL} is a natural tool for achieving this goal. The RL optimizes the actions of an agent (i.e., decision-maker) by interacting with the environment, and it has two salient features. First, it optimizes a {\it long-term} goal (rather than an instant utility function like in the previously mentioned QoE maximization works \cite{crosslayer1, crosslayer2, crosslayer3, crosslayer4, crosslayer5}). Second, RL enables
the agents to learn from their own experience through interactions and adaptation, without assuming prior knowledge of the external environment. As an extension of traditional RL, the deep reinforcement learning (DRL) frameworks~\cite{DQN,A3C} have recently become very popular. The idea is to use deep neural networks (DNN) to approximate the value or policy function in RL, so that complex and large state/value spaces can be parsimoniously represented by the DNNs. 
%However, it is still difficult to apply DRL into joint optimization for resource allocation and representation selection, which entails the DRL agent to simultaneously learn the two actions, namely, physical-layer transmission rate and application-layer chunk bitrate for each mobile DASH user. 

Indeed, there have been a few of recent works that apply DRL for streaming rate adaptation~\cite{pensieve, DQNDASH}. However, despite of being able to improve the long-term QoE performance, {\it a number of significant  challenges} emerge, as we list below:

\noindent  {\textsf{(1) Cross-Layer Optimization.}} It is generally difficult for the RL based scheme to account for the physical layer resource allocation. One possible reason is the time scale mismatch between the two tasks: The decision for physical layer transmission rate happens on the scale of milliseconds, %to keep pace with channel variation, 
while the selection for chunk bitrate is made in a much lower frequency (in seconds). In order to make RL based scheme adapt to the fast time scale physical layer decision making, the training complexity will be extremely high, due to the high-dimensional (possibly continuous) state and action space required to model physical layer objects such as channel state, transmit power, beamformers, user scheduling strategies, etc. 

%Another reason is that, even if a DRL-based model can be improve for such a prupose, the training complexity will be extremely high due to high-dimensional (possibly continuous) state and action space required to model physical layer resources such as transmit power, beamformers, use scheduling strategeis, etc. 

\noindent {\textsf{(2) Handling Multiple Users.}}  It is also difficult to model the behavior of multiple competing users under DRL framework, for a number of theoretical and practical reasons. First, the  interaction between competing users could be highly nonlinear and non-trivial (due to interference), therefore the usual ``stationarity" assumption on the environment is no longer valid; Second, in the presence of multiple users, the chunk requests of multiple users may occur at different moments, therefore  the decision steps for the chunk bitrate selection of each user are not synchronized.

% \noindent {\textsf{(3) Computational Complexity.}} 

%\noindent {\bf 2)} 

%This makes designing simple and trainable state and action spaces that are needed in DRL frameworks very challenging.
%In a nutshell, the asynchronous decision making and large state/action space impede the fully DRL-based cross-layer optimization.   
%where state space has high-dimensional even continuous values as inputs.
%Consequently, DRL-based dynamic power allocation~\cite{guodongning} as well as DRL-based rate adaptation schemes~\cite{pensieve}, \cite{DQNDASH} have been respectively studied  

\subsection{Contribution of This Work} 
In  this paper, we propose a novel two-level decision framework for multiuser adaptive video streaming in wireless networks. The main idea is to jointly design physical layer optimization-based scheme (performed at the base stations) and application layer DRL-based scheme (performed at the user terminals), so that a highly complex multi-user, cross-layer, time-varying video streaming problem can be decomposed into  relatively simple problems and solved effectively. Specifically, our two-level framework is explained below.

%In particular,  the bottom-layer wireless resource allocation is achieved by efficient and effective optimization-based methods, while DRL-based methods are independently performed at the user side for rate adaptation at the top layer.
%Specifically, i
In the physical layer, we formulate a quality-driven dynamic resource allocation problem, which optimizes transmit/receive beamformers to maximize a proportional fair utility function over some long-term average video quality measures. %of each user. %with proportional fairness consideration. %This problem is then converted to a weighted sum-quality maximization problem, which can be easily solved by extending the popular WMMSE algorithm\cite{WMMSE}. 
A {\it quality-driven dynamic resource allocation} (QDDRA) algorithm is then proposed, which is capable of determining the transmission rate for users in a multi-cell interfering broadcast channel. % by cross-cell cooperation in transmit beamformer design.

In the application layer, %to optimize the video streaming bitrates, 
we cast the adaptive video representation selection problem into a reinforcement learning task. This allows the mobile users to learn the desired chunk bitrate by interacting with changing wireless network environment, so that the long-term QoE is maximized. Particularly, we leverage a state-of-the-art DRL algorithm called asynchronous advantage actor-critic (A3C), which trains two neural networks, where the actor network is used to generate a policy (i.e, rate adaptation logic), together with a critic network to evaluate the learned policy. In addition, since low-motion video scenes (e.g., interviews or news) require less encoding bitrate to achieve a comparable QoE compared to high-motion situations (e.g., sport events or action movies), the video complexity is also taken into account when designing the state and reward.

Unlike the short-term cross-layer QoE maximization based works~\cite{crosslayer1, crosslayer2, crosslayer3, crosslayer4, crosslayer5}, which  do not consider long-term goals,
both QDDRA and DRL adopted in our framework are driven by maximizing long-term user service quality. Our approach is also different from the existing DRL frameworks that optimize solely over the application layer~\cite{pensieve,DQNDASH}, because our approach also takes the physical layer resource allocation into account.
Specifically, the QDDRA transmission rate allocation result is employed as a key state parameter for the subsequent DRL stage. 
Instead of defining extremely large state and action spaces for the cross-layer long-term optimization problem and casting everything into a single RL problem, the proposed method offers a simple yet effective way to integrate the tasks across the two layers---physical layer resource allocation and application layer long-term QoE maximization, and this simplicity makes implementation more reachable by practical systems. Extensive simulations and real-data experiments are conducted to showcase the effectiveness of our approach.

%The simulation results show that 
%decouple the complex problem into a centralized resource allocation controller and distributed rate adaptation
%Overall, the contributions of this paper can be summarized as follows.
%---
%1) We 
%
%2) We formulate
%
%3) We model
%
%4) We conduct extensive simulations under different system
%settings. The simulation results show that the proposed two-level decision framework 

The rest of this paper is organized as follows. %Section \uppercase\expandafter{\romannumeral2} reviews the related works in literature. 
In Section \uppercase\expandafter{\romannumeral2}, we introduce the framework and related system models. In Section \uppercase\expandafter{\romannumeral3}, we formulate a quality-driven dynamic resource allocation problem and present the corresponding solution. In Section \uppercase\expandafter{\romannumeral4}, we propose a DRL-based rate adaptation scheme that adapts the video quality to the time-varying wireless network to maximize the long-term user-perceived QoE. Section \uppercase\expandafter{\romannumeral5} presents experimental results, and evaluates the gains of the proposed algorithm compared to existing algorithms. The concluding remarks are given in Section\uppercase\expandafter{\romannumeral6}.

\noindent{\bf Notation.} Throughout the paper, we use capital bold-face letters to denote matrices while using the lower-case bold letters for vectors and small normal face for scalars. Moreover, we use the superscript $^H$ as the Hermitian transpose of a matrix and $\mathbf I$ as identity matrix. The notations $\mathbb E(\cdot)$ and $\det(\cdot)$ are used to represent the expectation and determinant operator respectively. In additional, $\mathbb C^{m\times n}$ is an $m$ by $n$ dimensional complex space, and $\mathcal {CN}(\cdot, \cdot)$ represents the complex distribution. 

%\section{Related Work}

\section{Framework and System models}
\subsection{Framework}

\begin{figure}[t]
  \centering
  \centerline{\epsfig{figure=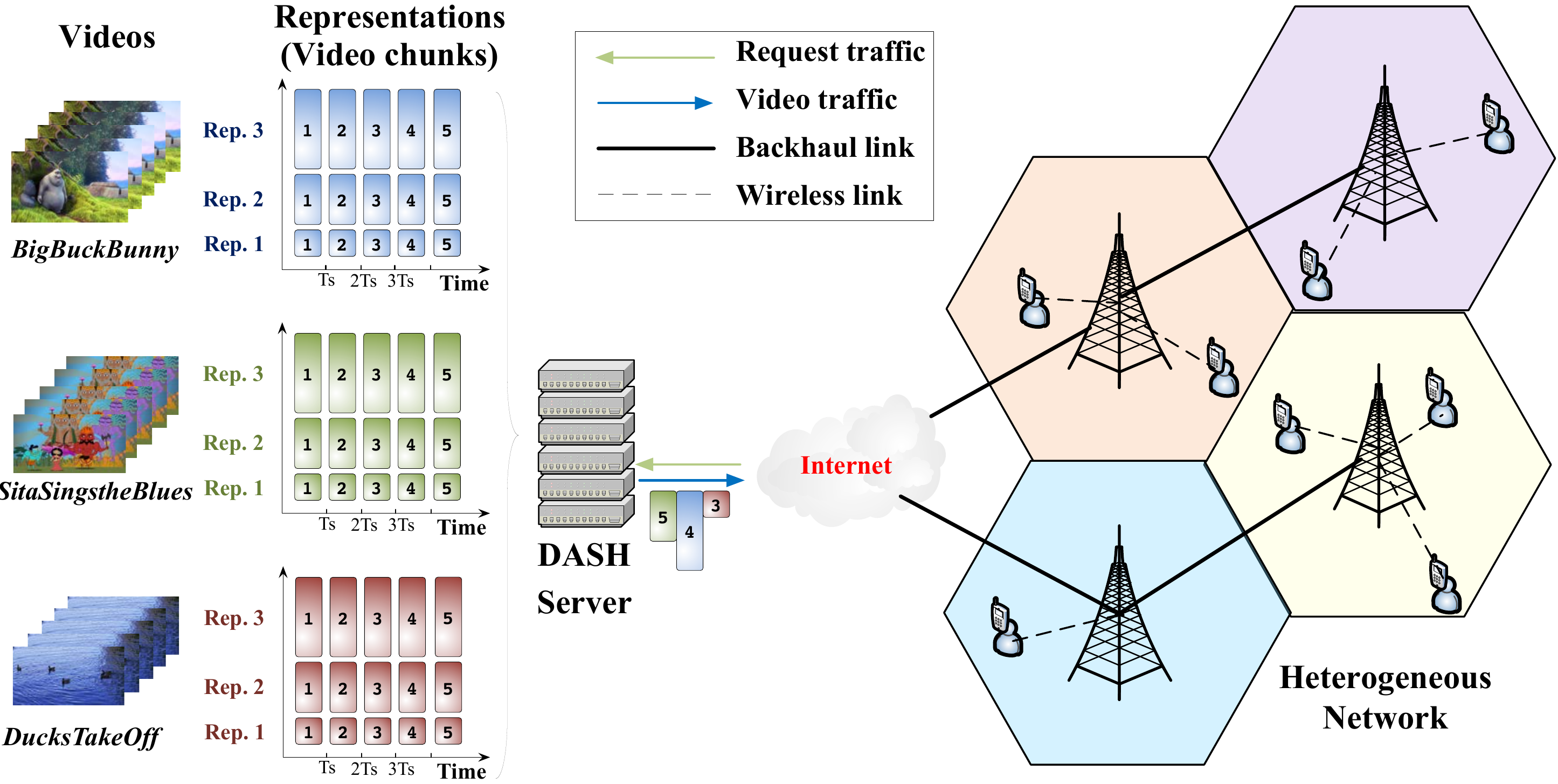,width=12cm}}
\caption{Example of the system considered in this work.}
  \label{fig:framework}
 \end{figure}

\begin{figure}[t]
  \centering
  \centerline{\epsfig{figure=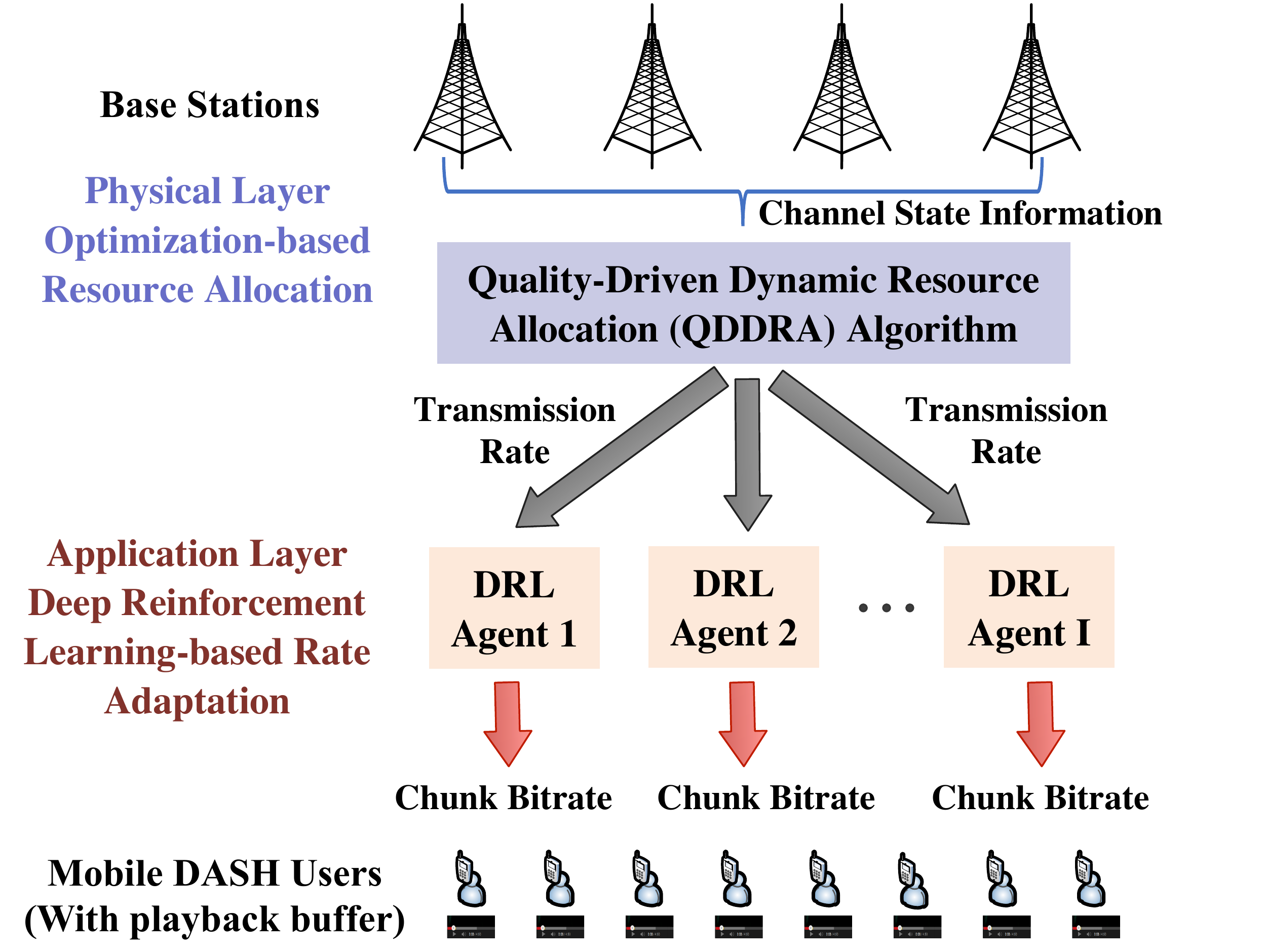,width=10cm}}
\caption{Illustration of the two-level decision framework proposed in this work.}
  \label{fig:framework2}
\end{figure}
As illustrated in Fig. \ref{fig:framework}, we consider a wireless adaptive streaming system consisting of a single DASH server and multiple DASH users located in a densely deployed heterogeneous network (HetNet). The server is connected to the HetNet through a high-speed backbone network. Suppose that the server stores $F$ video files denoted as $\mathcal F=\{1,2,\dots,F\}$, each of which is fragmented into small video chunks comprising $T_{\rm chunk}$ seconds of video. In other words, any video file $f\in\mathcal F$ can be treated as a set of consecutive video chunks indexed as $\{1,2,\dots,M_f\}$ where $M_f$ is the number of chunks for the file $f$. Each chunk is independently encoded into $L$ different representations (i.e., $L$ quality levels) with $\mathcal A_{f, m}$ being the available bitrate set for $m$-th chunk of the video file $f$. The information about available representations is recorded in the Media Presentation Description (MPD) file  that can be updated over time, and DASH users are able to obtain the MPD file by sending HTTP request. 

Furthermore, assume that there are $I$ mobile DASH users requesting video playback from the server, and they compete for the wireless resource in the cellular network. The video player of each user sends requests that specify the desired bitrate of chunk to be downloaded to the server at a time, and each chunk must be completely downloaded into the playback buffer before being decoded and watched. The channel states between users and base stations could frequently change due to user mobility and channel fading. Here, we consider a time-slotted system with each time slot being  $T_{\rm slot}$ seconds long, and channel states are stable within the duration of a time slot. The time-slotted system is widely used in wireless communications \cite{Mihaela}, and the specified length of one time slot depends on how fast the channel changes. Moreover, we assume that the channel state information (CSI) can be precisely measured at each receiver and promptly transmitted to a central controller that is responsible for wireless resource allocation \cite{Multicell-Gesbert}. According to the cross-cell CSI at each time slot $t$, the controller will determine the transmit beamforming vector for each base station-user pair to achieve high system performance while ensuring user fairness in terms of received video quality. On the other hand, each user will dynamically adapt the chunk bitrate to the time-varying wireless resources allocated by the controller such that its long-term quality-of-experience (QoE) is maximized. The main symbols are summarized in Table \ref{tab:notations}.  

As shown in Fig. \ref{fig:framework2}, we remark that in this framework, physical layer coordination is used to improve user fairness and overall physical layer resource utilization. 
Long-term user QoE is locally optimized by each user at the second stage.
Hence, there is no centralized computation or synchronization required on the application layer. This is very different from the existing short-term cross-layer QoE optimization works in ~\cite{crosslayer1, crosslayer2, crosslayer3, crosslayer4, crosslayer5}. Arguably,  our framework better fits the user-dependent nature of video streaming services.

%$T_{slot}$ & The length of a time slot in seconds.
\begin{table*}[t]
\caption{Main Notations Used in This Work}\footnotesize
\centering
\begin{tabular}{|c|m{5.1cm}<{\centering}|c|m{5.3cm}<{\centering}|}
 	\hline
 	\multicolumn{4}{|c|}{\textbf{The Wireless Model}}\\
 	\hline 
	\textbf{Symbol}&\textbf{Description}&\textbf{Symbol}&\textbf{Description}\\
	\hline
	 $\mathcal K=\{1,2,\dots,K\}$ & The set of $K$ cells/base stations. & $i_k$ & The $i$-th user in cell $k$.\\
	\hline 
	$\mathcal I_k=\{1_k,2_k,\dots,I_k\}$ &	The set of $I_k$ users located in cell $k$. & $\mathcal I = \bigcup^K_{k=1}\mathcal I_k$ & The set of all users. \\
	\hline
	$N^{\rm ta}_k$ & The number of transmit antenna of base station $k\in\mathcal K$. & $N^{\rm ra}_{i_k}$ & The number of receive antenna of mobile user $i_k\in\mathcal I$.\\
	\hline
	$\mathbf v_{i_k}^t$ & The transmit beamformer of base station $k$ to user $i_k\in\mathcal I_k$ at time slot $t$. & 	$\mathbf H_{i_k, k}^t$ & The channel gain from base station $k$ to user $i_k\in\mathcal I$  at time slot $t$.\\
	\hline
	$\mathbf u_{i_k}^t$ & The receive beamformer of user $i_k\in\mathcal I$ at time slot $t$. &
	$R_{i_k}^t $ & The achievable data transmission rate for user $i_k\in\mathcal I$ at time slot $t$.\\
	\hline
	\multicolumn{4}{|c|}{\textbf{QoE Model}}\\
	\hline
	$\mathcal F=\{1,2,\dots,F\}$ & The set of $F$ video streams.&  $T_{\rm chunk}$ & The length of a chunk in seconds.\\
	\hline
   $\{1,2,\dots,M_f\}$ & The chunk index of video $f\in\mathcal F$ with $M_f$ video chunks.& $\mathcal A_{f, m}$ & The available bitrate set of the $m$-th chunk of video $f\in\mathcal F$.\\ 
    \hline
    $a_m$ & The selected bitrate of chunk $m$. & $q_m$  & The perceived video quality of chunk $m$.\\
    \hline
    $\mathbf z_m$ & The content complexity of chunk $m$. & $d_m$ & The download time of the chunk $m$.\\
    \hline
    $C_m$ & The experienced average throughput while downloading chunk $m$. & $b_m$ & The buffer occupancy when the player starts downloading chunk $m$.\\
    \hline 
\end{tabular}
\label{tab:notations}
\end{table*}

\subsection{Wireless Network Model}
To formulate the physical layer resource allocation problem,
we consider a HetNet with $K$ cells, modeled by MIMO interfering broadcast channel (IBC). 
Note that considering multiple antennas on the mobile side is well-motivated, since the next-generation mobile phones are expected to carry relatively large antenna arrays.
There is a  single base station $k\in\mathcal K=\{1,2,\dots,K\}$ within cell $k\in\mathcal K$, and it is equipped with $N^{\rm ta}_k$ transmit antennas and sends video data to $I_k$ users located in different areas of the cell $k$. Let $i_k$ denote the $i$-th user in cell $k$ who has $N^{\rm ra}_{i_k}$ receive antennas, and the set of all users is represented as $\mathcal I = \bigcup^K_{k=1}\mathcal I_k$ where $\mathcal I_k=\{1_k,2_k,\dots,I_k\}$ is the set of users located in cell $k$. Denote by $\mathbf v_{i_k}^t\in\mathbb C^{N^{\rm ta}_k}$ the transmit beamformer of base station $k\in\mathcal K$ to user $i_k\in\mathcal I_k$ and $o_{i_k}^t\in\mathbb C$ the transmitted data symbol with unit variance at time slot $t$, then the transmitted signal for user $i_k$ can be characterized as:   
\begin{equation}\label{equ:transmittedsignal}
\mathbf x_{i_k}^t = \mathbf v_{i_k}^to_{i_k}^t.
\end{equation}
Taking into account a linear channel model, the corresponding received signal of user $i_k$ at time slot $t$ can be expressed as:
\begin{equation}\label{equ:receivedsignal}
\begin{split}
\mathbf y_{i_k}^t&=\underbrace{\mathbf H_{i_k, k}^t\mathbf x_{i_k}^t}_{\textrm{intended signal}} + \underbrace{\sum_{j_k\in\mathcal I_k\backslash\{i_k\}}\mathbf H_{i_k, k}^t\mathbf x_{j_k}^t}_{\textrm{intracell interference}} \\ &+ \underbrace{\sum_{j_l\in\mathcal I\backslash\mathcal I_k}\mathbf H_{i_k, l}^t\mathbf x_{j_l}^t}_{\textrm{intercell interference}} +\underbrace{n_{i_k}^t}_{\textrm{noise}}, \quad\quad \forall i_k\in\mathcal I,  
\end{split}
\end{equation}
where $\mathbf H_{i_k, k}^t\in\mathbb C^{N^{\rm ra}_{i_k}\times N^{\rm ta}_k}$ represents the downlink channel gain from the base station $k$ to users $i_k$, and $\mathbf n_{i_k}^t\in\mathbb C^{N^{\rm ra}_{i_k}\times 1}$ denotes the additive white Gaussian noise (AWGN) at user $i_k$ with probability distribution $\mathcal {CN}(0, \sigma_{i_k}^2\mathbf I_{N^{\rm ra}_{i_k}})$. Further, assume that each user treats the interference as noise and adopts linear receive beamforming strategy, then the estimated data symbol at user $i_k$ can be written as:
\begin{equation}\label{equ:receivedsymbol}
	\widehat{o_{i_k}^t}= (\mathbf u_{i_k}^t)^H\mathbf y_{i_k}^t
\end{equation}
with $\mathbf u_{i_k}^t\in\mathbb C^{N^{\rm ra}_{i_k}\times 1}$ being the receive beamformer of user $i_k$ at time slot $t$. Further, the signal-to-interference-plus-noise ratio (SINR) at the receiver $i_k$ can be written as:
%\textrm{SINR}_{i_k}^t=\frac{||\mathbf H_{i_k, k}^t\mathbf v_{i_k}^t||^2}{\sum_{(l, j)\neq(k, i)}||\mathbf H_{i_k, l}^t\mathbf v_{j_l}^t||^2+\sigma_{i_k}^2}, \quad \forall i_k\in\mathcal I.
\begin{equation}\label{equ:SINR}
\textrm{SINR}_{i_k}^t=\mathbf H_{i_k, k}^t\mathbf v_{i_k}^t{(\mathbf v_{i_k}^t)}^H{(\mathbf H_{i_k, k}^t)}^H\mathbf \Theta_{i_k}^{-1},\quad \forall i_k\in\mathcal I,
\end{equation}
where $$\mathbf\Theta_{i_k}=\sum_{(l, j)\neq(k, i)}\mathbf H_{i_k, l}^t\mathbf v_{j_l}^t{(\mathbf v_{j_l}^t)}^H{(\mathbf H_{i_k, l}^t)}^H+\sigma_{i_k}^2\mathbf I_{N^{\rm ra}_{i_k}}.$$ Accordingly, the achievable data transmission rate for user $i_k$ is 
\begin{equation}\label{equ:transrate}
R_{i_k}^t = B\log_2\det\left(\mathbf I_{N^{\rm ra}_{i_k}}+\frac{\textrm{SINR}_{i_k}^t}{\Gamma}\right), \quad \forall i_k\in\mathcal I,
\end{equation}
where $B$ represents the channel bandwidth and $\Gamma$ is the SNR gap depending on the modulation scheme.

\subsection{QoE Model}
%The wireless models given above are used for physical layer resource allocation. 
To appropriately carry out the application layer rate adaptation, it is critical to identify a well-defined QoE model which measures use satisfaction. According to~\cite{survey}, the QoE of DASH user is greatly influenced by three key factors, namely, the received video quality, the quality variations and the frequency of the rebuffering events (i.e., scenarios where the playback buffer is empty and the user needs to wait until the next chunk is downloaded). Suppose that a user $i_k\in\mathcal I$ requests the playback for the video file $f\in\mathcal F$ and downloads the video chunks with the corresponding representation in turn. We denote by\footnote{All variables defined in this subsection are associate with a user $i_k\in\mathcal I$ , though we omit the subscript $i_k$ for brevity.} $a_m\in\mathcal A_{f, m}$ the selected bitrate for the $m$-th chunk and $q_m$ the video quality perceived by the user about the chunk. The relationship between video bitrate and quality can be depicted by a parametric rate-quality function that maps the encoding bitrate to some quality metrics such as Peak Signal-to-noise ratio (PSNR), Structural Similarity Index (SSIM) and mean opinion score (MOS). Therefore, the quality of the chunk $m$ with representation bitrate being $a_m$ can be expressed as: 
\begin{equation}\label{equ:quality}
	q_m=g(a_m;\mathbf z_m),
\end{equation}
where $\mathbf z_m$ is a content-dependent parameter vector that indicates the complexity of the chunk $m$. The function $g(\cdot):\mathbb R_+\to \mathbb R_+$ is a continuous, invertible and strictly increasing function of the video bitrate, the use of which indicates that a video chunk with a higher bitrate will produce a higher video quality. %In general, a logarithmic function. 
One typical example of such function is \cite{crosslayer2,crosslayer3}:
\begin{equation}\label{equ:quality1}
	q_m=z_{m,1}\log (z_{m,2}a_m+z_{m,3}),
\end{equation}
where the parameters $z_{m,1}$, $z_{m,2}$ and $z_{m,3}$ can be estimated by curve-fitting over some empirically discrete points~\cite{ratedistortion}.

The download time $d_m$ of the chunk $m$ depends on its size in bits as well as the allocated physical layer downlink transmission rate. Assume that the video player starts to download chunk $m$ at time $t_m$,  then the experienced average throughput is defined as \cite{MPC}
\begin{equation}\label{equ:averagespeed}
	C_m=\frac{1}{t_{m+1}-t_m}\int_{t_m}^{t_{m+1}} R^t dt,
\end{equation}
where $R^t$ is transmission rate for the user defined in Eq. \eqref{equ:transrate}. Here, $t$ denotes a continous time index, and $R^t$ can be treated as a piecewise function which changes its  corresponding rate value every $T_{\rm slot}$ seconds. For simplicity, we assume that the player will immediately request the next chunk once the the current chunk is entirely received. Hence, the time $t_{m+1}$ to request $(m$+1)-th chunk is equal to the time when the chunk $m$ is downloaded. If the buffer is full, then the player will wait for a period of time denoted as $\Delta t_m$ before requesting the next chunk. Accordingly, the integral upper limit and the denominator become $t_{m+1}-\Delta t_m$ and $t_{m+1}-\Delta t_m-t_m$ respectively. Then, the download time can be derived as $d_m=\tau_m(a_m)/C_m$, where $\tau_m(a_m)$ denotes the size of chunk $m$ encoded at bitrate $a_m$. %(a higher bitrate implies a larger chunk size). 
The downloaded video chunks are stored in the user's playout buffer. Let $b_m\in[0, b_{\rm max}]$ denote the buffer occupancy (measured in seconds) when the video player attempts to request chunk $m$. The buffer size $b_{\rm max}$ depends on the storage limitation of the display device. A rebuffering event occurs when the buffer becomes empty. That is, the download time $d_m$ is larger than the buffer level $b_m$, so that no video content can playback before the next chunk is completely downloaded. Accordingly, the rebuffering time can be formulated as:
\begin{equation}\label{equ:rebuffering}
\phi_m = (d_m-b_m)_+,
\end{equation}
where the notation $(x)_+=\max\{x,0\}$. In addition, we can derive the evolution of the buffer occupancy, that is, the buffer level at time $t_{m+1}$ as $b_{m+1}=(b_m-d_m)_++T_{\rm chunk}$. Similarly, if the buffer becomes full after downloading the chunk $m$, then we have the buffer level at time $t_{m+1}$ being $b_{m+1}=((b_m-d_m)_++T_{chunk}-\Delta t_m)_+$ with $\Delta t_m$ being the waiting time.

To achieve the efficiency of the adaptive streaming system, the user should watch the highest possible video quality based on its available channel capacity. Meanwhile, frequent quality switches and rebuffering events should be avoided to guarantee smooth and stall-free playback. Therefore, we can define the user-perceived QoE of the chunk $m$ as
\begin{equation}\label{equ:QOE}
	QoE_m=q_m-\lambda|q_m-q_{m-1}|-\rho\phi_m,
\end{equation}
where $\lambda$ and $\rho$ are non-negative parameters used to trade off the instantaneous video quality, quality fluctuations and rebuffering events in the QoE evaluation. 
%-\delta (b_{\textrm{th}}-b_{m+1})_+
%The last term is used to further reduce the risk of rebuffering events, and there is a penalty when the buffer level is less than a given threshold $b_{\textrm{th}}$ that can be regarded as a ``safe'' buffer level.

\section{Quality-Driven Dynamic Resource Allocation}
Considering that the wireless resource allocation at the physical layer has a significant influence on the ultimate performance of adaptive streaming system, in this section, we formulate a proportional fairness maximization problem in terms of long-term average video quality for multiple DASH users in HetNet, and design a Quality-Driven Dynamic Resource Allocation (QDDRA) algorithm  accordingly.
\subsection{Problem Formulation}
In accordance with the QoE model defined in Section \uppercase\expandafter{\romannumeral3}-C, it can be seen that the user-perceived QoE in the application layer is closely coupled with the allocated transmission rate in the physical layer. Therefore, we first exploit the parametric rate-quality function to map the low-level transmission rate into high-level video quality, and consider a quality-driven resource allocation problem that maximizes the sum-quality  of all users at each time slot. To be specific,  the video quality of user $i_k\in\mathcal I$ at time slot $t$ is given by:
\begin{equation}\label{equ:videoquality}
\begin{split}
q_{i_k}^t&=g_{i_k}(R_{i_k}^t;\mathbf z_{i_k}^t)\\&=z_{i_k,1}^t\log (z_{i_k,2}^tR_{i_k}^t+z_{i_k,3}^t),	
\end{split}
\end{equation}
where the parameter vectors $\mathbf z_{i_k}^t$ is the complexity of video viewed by the user $i_k$ at the current time slot that can be directly extracted from the MPD file. The variable $R_{i_k}^t$ computed by Eq. \eqref{equ:transrate} is the achievable transmission rate for user $i_k$, which is determined by the beamforming weights at time slot $t$. 

The sum-quality maximization resource allocation problem can be summarized as follows. At each time slot $t$, given the channel state information, namely, the set of channel gain matrix $\{\mathbf H_{i_k, k}^t\}_{k\in\mathcal K, i_k\in\mathcal I_k}$ between base stations and mobile users as well as user-viewed video complexity information, the optimization objective is to find the transmit beamformers $\mathbf v^t$ for all base station-user pairs such that the total users' video qualities is maximized, subject to the transmit power constraints at the base stations\footnote{The notation $\mathbf v^t$ is short for $ \{\mathbf v^t_{i_k}\}_{i_k\in\mathcal I}$, which denotes all variables $\mathbf v_{i_k}^t$ with $i_k\in\mathcal I$. The short notations $\mathbf u^t=\{\mathbf u_{i_k}^t\}_{i_k\in\mathcal I}, w^t=\{w_{i_k}^t\}_{i_k\in\mathcal I}$ are defined similarly.}. Mathematically, such a problem can be formulated as:
\begin{equation}\label{equ:P1}
\begin{split}
&\textrm{\textbf{P1:}} \quad \max_{\mathbf v^t} \sum_{i_k\in\mathcal I}	q_{i_k}^t\\ &\textrm{s.t.} \quad \sum_{i_k\in\mathcal I_k}||\mathbf v_{i_k}^t||^2_2\leq P_{k}, \quad \forall k\in\mathcal K,
\end{split}
\end{equation}
where $P_k$ denotes the power budget of base station $k$.

Furthermore, due to user mobility and wireless channel fading, the downlink channel conditions between base stations and mobile users dynamically change over time. Therefore we further investigate a quality-driven dynamic resource allocation problem that maximizes the sum of long-term average quality of each user over a period of time. Meanwhile, proportional fairness is considered to achieve quality fairness among users. Hence, the quality-driven dynamic resource allocation problem for multiuser adaptive video streaming transmission in heterogeneous network can be formulated as:
\begin{equation}\label{equ:P2}
\begin{split}
&\textrm{\textbf{P2:}} \quad \max_{\mathbf v^t} \sum_{i_k\in\mathcal I}	\log Q_{i_k}^t\\ &\textrm{s.t.} \quad \sum_{i_k\in\mathcal I_k}||\mathbf v_{i_k}^t||^2_2\leq P_{k}, \quad \forall k\in\mathcal K,
\end{split}
\end{equation}
where $Q_{i_k}^t$ represents the long-term average quality of user $i_k$ up to time slot $t$, i.e., 
\begin{equation}\label{equ:averagequality}
	Q_{i_k}^t=\beta q_{i_k}^t+(1-\beta)Q_{i_k}^{t-1},
\end{equation}
where $\beta\in(0,1]$ is used to control the impact of average video quality obtained in the pervious time slots. For example, using exponential averaging, we have $\beta=T_{\rm slot}/T_{\rm win}$ where $T_{\rm slot}$ and $T_{\rm win}$ are the length of one time slot and the predefined averaging window size respectively. Likewise, $q_{i_k}^t$ denotes the video quality of user $i_k$ at time slot $t$ that is determined by the current beamforming vectors $\mathbf v^t$.
 
In practice, proportional fairness maximization problem can be approximately implemented using a weighted sum maximization problem~\cite{logfair}, \cite{logfair2}. Consequently, problem \textbf{P2} can be converted to the following weighted sum-quality maximization problem:
\begin{equation}\label{equ:P3}
\begin{split}
&\textrm{\textbf{P3:}} \quad \max_{\mathbf v^t} \sum_{i_k\in\mathcal I}	\alpha_{i_k}^t q_{i_k}^t \\ &\textrm{s.t.} \quad \sum_{i_k\in\mathcal I_k}||\mathbf v_{i_k}^t||^2_2\leq P_{k}, \quad \forall k\in\mathcal K
\end{split}
\end{equation}
with $\alpha_{i_k}^t=1/Q_{i_k}^{t-1}$. The solution for problem \textbf{P3} is similar to the problem \textbf{P1} since \textbf{P3} just adds a weight factor to the objective function. Albeit the weights are changing over time, they are constants at each time slot. Besides, the solution for the problem \textbf{P1} can be derived by extending the popular WMMSE algorithm~\cite{WMMSE}. In this way, we simplify the quality-driven dynamic resource allocation problem into an easily-solved weighted sum-quality maximization problem.

\subsection{Algorithm Design}
Let $e_{i_k}^t\in\mathbb R_+$ be the mean-square error (MSE) received by user $i_k$ at time slot $t$ using the well-known minimizing sum-MSE (MMSE) receiver. As shown in Appendix A of~\cite{WMMSE}, the relation between the MSE $e_{i_k}^t$ and transmission rate $R_{i_k}^t$ can be expressed as:  
\begin{equation}\label{equ:R-E}
R_{i_k}^t=\log (e_{i_k}^t)^{-1}.
\end{equation}
Therefore, we can reformulate the problem \textrm{\textbf{P1}} as the following sum-MSE cost minimization problem: 
\begin{equation}\label{equ:P4}
\begin{split}
&\textrm{\textbf{P4:}} \quad \min_{\mathbf v^t, \mathbf u^t} \sum_{i_k\in\mathcal I}	c_{i_k}(e_{i_k}^t) \\ &\textrm{s.t.} \quad \sum_{i_k\in\mathcal I_k}||\mathbf v_{i_k}^t||^2_2\leq P_{k}, \quad \forall k\in\mathcal K,
\end{split}
\end{equation}
where $c_{i_k}(\cdot)$ is the cost function of receiver $i_k$ that can be derived by substituting the $R_{i_k}$ in Eq. \eqref{equ:videoquality} with Eq. \eqref{equ:R-E} along with a minus sign since the optimization problem has changed into minimizing the objective function. That is, we have 
\begin{equation}\label{equ:cost}
\begin{split}
	c_{i_k}(e_{i_k}^t)&=-g_{i_k}(-\log e_{i_k}^t;\mathbf z_{i_k}^t)\\&=-z_{i_k,1}^t\log(-z_{i_k,2}^t\log e_{i_k}^t+z_{i_k,3}^t).
\end{split}
\end{equation}

Further, by introducing auxiliary weight variables $\{w_{i_k}^t\}_{i_k\in\mathcal I}$, we can define the following weighted sum-MSE minimization problem:
\begin{equation}\label{equ:P5}
\begin{split}
&\textrm{\textbf{P5:}} \quad \min_{\mathbf v^t, \mathbf u^t, w^t} \sum_{i_k\in\mathcal I}	 w_{i_k}^te_{i_k}^t+c_{i_k}(\Upsilon_{i_k}(w_{i_k}^t))-w_{i_k}^t\Upsilon_{i_k}(w_{i_k}^t) \\ &\textrm{s.t.} \quad \sum_{i_k\in\mathcal I_k}||\mathbf v_{i_k}^t||^2_2\leq P_{k}, \quad \forall k\in\mathcal K,
\end{split}
\end{equation}
where $\Upsilon_{i_k}(\cdot)$ is the inverse function of the derivate function of the function $c_{i_k}(\cdot)$. Based on the Theorem 2 in~\cite{WMMSE}, problem $\textbf{P5}$ is equivalent to the problem $\textbf{P4}$ in the sense that they have the same global optimal solution, if $c_{i_k}(\cdot)$ is a strictly concave function of $e_{i_k}^t$ for all ${i_k}$. Moreover, given the transmit-receive beamformer pair $\{\mathbf v_{i_k}^t, \mathbf u_{i_k}^t\}$ for each user $i_k$, the optimal weight $w_{i_k}^t$ for problem $\textbf{P5}$ is given by $w_{i_k}^*=c_{i_k}^\prime (e_{i_k}^t)$ where $c_{i_k}^\prime(\cdot)$ is the derivative of the function $c_{i_k}(\cdot)$. Obviously, the cost function $c_{i_k}(\cdot)$ in our problem is strictly concave. Hence, similar to the WMMSE algorithm \cite{WMMSE} that leverages the block coordinate descent method to solve the sum-MSE minimization problem, we can iteratively update one of the three variables by fixing the remaining two variables to solve problem \textbf{P5}. The quality-driven dynamic resource allocation (QDDRA) algorithm for multiple mobile DASH users in HetNet is proposed in Algorithm \ref{alg}.

\begin{algorithm}[t]
 \renewcommand{\baselinestretch}{1.0}
 \renewcommand{\arraystretch}{1.0}
 \small
 \caption{QDDRA algorithm.}\label{alg}
 \begin{algorithmic}[1]
 \STATE \textbf{At} the current time slot $t$ \textbf{do}
 \STATE \quad\textbf{Input:} $\alpha_{i_k}^t=1/Q_{i_k}^{t-1}$, $\mathbf z_{i_k}^t$, $\mathbf H_{i_k, l}^t$;$\forall~l\in\mathcal K, i_k\in\mathcal I$;\\
 \STATE \quad\textbf{Initialize} $\mathbf v_{i_k}^t$ such that $||\mathbf v_{i_k}^t||^2_2= \frac{P_k}{I_k}$, $\forall~i_k\in\mathcal I$;\\
 \STATE \quad\textbf{Repeat} for each user $i_k\in\mathcal I$\\
% \STATE \quad \quad $w_{i_k}^\prime\gets w_{i_k}$,  $\forall~i_k\in\mathcal I$;\\
 \STATE \quad \quad $\mathbf u_{i_k}^t\gets \left(\sum\limits_{(l, j)}\mathbf H_{i_k, l}^t\mathbf v_{j_l}^t(\mathbf v_{j_l}^t)^H(\mathbf H_{i_k, l}^t)^H+\sigma_{i_k}^2\mathbf I\right)^{-1}\mathbf H_{i_k, k}^t\mathbf v_{i_k}^t$; \\
  \STATE \quad \quad $w_{i_k}^t\gets c_{i_k}^\prime (e_{i_k}^t)|_{e_{i_k}^t=1-(\mathbf u_{i_k}^t)^H\mathbf H_{i_k, k}^t\mathbf v_{i_k}^t}$;\\
 \STATE \quad \quad $\mathbf v_{i_k}^t\!\!\!\gets\!\!\!\alpha_{i_k}^t \!\!\left(\!\sum\limits_{(l, j)}\!\!\!\alpha_{j_l}^t\! (\mathbf H_{j_l, k}^t)^H\mathbf u_{j_l}^t \!w_{j_l}^t \!(\mathbf u_{j_l}^t)^H \!\mathbf H_{j_l, k}^t\!\!+\!\!\mu_k^*\mathbf I\right)^{-1}\!\!\!\!\!\!\!(\mathbf H_{i_k, k}^t)^H \!\mathbf u_{i_k}^t \!w_{i_k}^t$\!;\\
  \STATE \quad\textbf{Until} Some stopping criteria is met.\\%$\left|\sum_{j=1}^K \log\left(w_j^t\right){-}\sum_{j=1}^K \log\left(w_j^{t-1}\right)\right|{\leq} \epsilon$;\\
  \STATE \quad \textbf{Compute} $R_{i_k}^t$, $q_{i_k}^t$, $Q_{i_k}^t$ based on Eq. \eqref{equ:transrate}\eqref{equ:videoquality}\eqref{equ:averagequality} respectively.\\
 \STATE \quad \textbf{Output:} $R_{i_k}^t$, $q_{i_k}^t$, $Q_{i_k}^t$, $\forall~i_k\in\mathcal I$;\\
 \STATE\textbf{Update} $t \gets t+1$\\
 \end{algorithmic}
\end{algorithm}

\section{DRL-Based Rate Adaptation}
Once the transmission rates for DASH users are determined on the physical layer, users can separately select and switch their video representations at each chunk boundary such that their respective long-term QoEs are maximized. In this section, we first formulate the optimal representation (chunk bitrate) selection problem at the user side into a RL problem. In particular, the video complexity is taken into account in designing state and reward. Then, we employ the asynchronous advantage actor-critic (A3C), a popular DRL algorithm to train actor and critic neural network, to  approximate the policy function and value function respectively. As a consequence, the learned policy is able to select the best action (representation) based on the system state (measured throughput, buffer occupancy, video complexity, etc.) to maximize the long-term reward (QoE).
\subsection{Problem Formulation}
For any mobile DASH user $i_k\in\mathcal I$ who is assumed to request the playback of video file $f\in \mathcal F$, the aim is to find the optimal representation vector $\mathbf a^*\in\{[a_1,\dots,a_{M_f}]|a_m\in\mathcal A_{f, m}\}$ that maximizes the aggregated QoE of all chunks contained in the video $f$. More formally, the optimal representation (chunk bitrate) selection problem performed at the user side can be formulated as:
\begin{equation}\label{equ:maxlongtermQoE}
	\textrm{\textbf{P6:}} ~ \max_{\mathbf a}\sum_{m=1}^{M_f}QoE_m,
\end{equation}
where $QoE_m$ is the user-perceived QoE of the chunk $m$ defined in Eq. \eqref{equ:QOE}. 
%While the finite-horizon optimal control problem \textbf{P6} is well-defined for the QoE maximization in rate adaptation, it requires the exact throughput trace $\{C_m\}_{m\in\mathcal M_f}$ and chunk complexity trace $\{\mathbf z_m\}_{m\in\mathcal M_f}$ as inputs, which in practice can only be estimated based on throughput/complexity predictors. However, it is difficult to capture the system dynamics (e.g., how the network throughput/chunk complexity vary over time) since the experienced throughput of each user for downloading individual chunk is intricately influenced by the time-varying wireless channel conditions and the wireless resources actually obtained by the user at each time slot. 
Note that optimizing \textrm{\textbf{P6}} using deterministic one-shot optimization techniques is not easy, since the $QoE_\ell$'s for $\ell>m$ are affected by many factors, including actions taken by the user at chunk $m$, the feedback from the external environment, and the link quality in the physical layer---which are all hard to know in advance.
Hence, we consider a RL framework where an agent (e.g, the mobile DASH user) learns the best action (i.e., the best bitrate of chunk to be downloaded) to achieve the anticipated goal (i.e., maximizing the long-term QoE) from the interaction with the environment.

 The agent-environment interaction process in RL can be described as follows. At each decision step $m$, the agent observes the state $s_m$ and selects an action $a_m$ from the set of possible actions $\mathcal A(s_m)$ according to its policy $\pi(a_m|s_m)$ that specifies the probability of selecting $a_m$ in state $s_m$. As a consequence of the action, the agent receives a scalar reward $r_m$ and observes the next state $s_{m+1}$. The goal of the agent is to maximize the expected return (i.e., the total accumulated reward) $\mathbb E~[\sum_{m=0}^{\infty} \gamma^m r_{m}]$, where $\gamma\in [0, 1]$ is a discount factor that determines the present value of future rewards. The value function $V^{\pi_\theta}(s_m)$ (or state-action value function $Q^{\pi_\theta}(s_m, a_m)$) represents the expected return when starting from state $s_m$ (or starting from state $s_m$ and taking action $a_m$), and thereafter follows policy $\pi$. Accordingly, we define the state, action and reward for the rate adaptation process as:

\textbf{1) State:} The system state at the time when the video player starts to request the download of the chunk $m$ is defined as $s_m=(\overline{\mathbf C}_m, \overline{\mathbf d}_m, \mathbf z_m, \bm \tau_m, b_m, \omega_m, \delta_m)$, which is characterized by the measured network throughput experienced by the previous $n$ video chunks, $\overline{\mathbf C}_m=[C_{m-n},\dots,C_{m-1}]$, where each $C_m$ is given in \eqref{equ:averagespeed}, and the rates are computed by the QDDRA algorithm developed in Section III; the download time of the past $n$ video chunks, $\overline{\mathbf d}_m=[d_{m-n},\dots,d_{m-1}]$; the complexity of the chunk $m$, $\mathbf z_m$; the available sizes of the chunk $m$, $\bm \tau_m$; the current buffer occupancy, $b_m$; the number of remainder chunks in the video, $\omega_m$; and the video quality of the last downloaded chunk, $\delta_m=q_{m-1}$. \textbf{2) Action:} The action $a_m$ corresponds to the selected bitrate for the chunk $m$. \textbf{3) Reward:} The scalar reward is an available immediate feedback from the environment when the agent takes an action, here we consider the QoE of chunk $m$ as reward, that is, $r_m=QoE_m$. Compared to problem \textbf{P6}, the aim is changed to find the optimal policy $\pi^*:\mathcal S \times \mathcal A\rightarrow [0,1]$ such that the expected long-term (discounted) QoE is maximized. Thus, the optimal representation (chunk bitrate) selection problem performed at the user side is rewritten as\footnote{Although that users watch adaptive video streaming is an episodic task which has a terminal state when requesting the last chunk in video file, for convenience of expression, we adopt the unified notation (i.e., taking infinity as upper limit) for episodic and continuing tasks by setting episode termination to be the entering of a special absorbing state that transitions only to itself and that generates only rewards of zero. Correspondingly, we also adjust the chunk index to start from zero.}:
\begin{equation}\label{equ:goal}
		\textrm{\textbf{P7:}}~ \max_{\pi} ~ \mathbb E_{\pi}[\sum_{m=0}^{\infty} \gamma^{m}r_m].
\end{equation}
%\begin{equation}\label{equ:transition}
%\begin{split}
%		p(s_{m+1}|s_m, a_m)&=p(\overline{\mathbf C}_{m+1}|\overline{\mathbf C}_m)\times p(\overline{\mathbf d}_{m+1}|\overline{\mathbf d}_m, a_m)\times p(\mathbf z_{m+1}|\mathbf z_m)\\
%		&\times p(\bm \tau_{m+1}|\bm \tau_m)\times p(b_{m+1}|b_m, a_m)\\ &\times p(\omega_{m+1}|\omega_m)\times p(l_{m+1}|\mathbf z_m, a_m)
%\end{split}
%\end{equation}

\subsection{A3C Algorithm}
\begin{figure}[t]
  \centering
  \centerline{\epsfig{figure=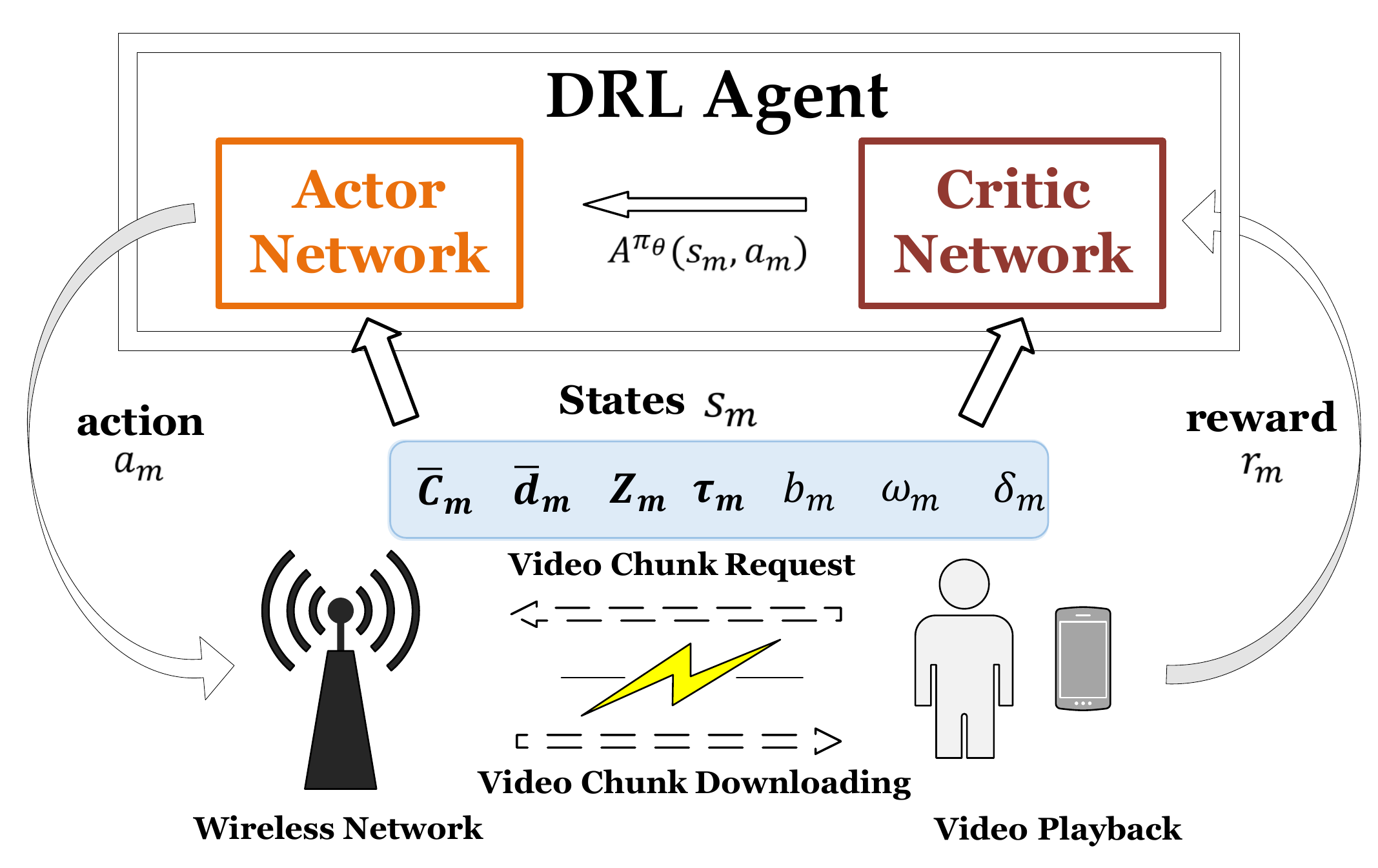,width=10cm}}
\caption{Framework of DRL-based rate adaptation.}
 \label{fig:A3C}
\end{figure}
Since some of system state parameters described in last subsection are continuous real numbers, which inevitably leads to a very large state space.
Hence, we leverage a neural network with parameters $\theta$ to approximate the policy $\pi$, then we can represent the policy as $\pi_\theta(a_m|s_m;\theta)$. Besides, in order to solve the problem \textbf{P7}, we take into account a policy gradient method that directly learns the parametrized policy $\pi_\theta$ and updates the policy parameters $\theta$ by performing gradient ascent on the expected total reward. Specifically, as illustrated in Fig. \ref{fig:A3C}, we propose to employ the asynchronous advantage actor-critic (A3C) algorithm \cite{A3C}. There, an actor neural network with a softmax output is used to maintain a policy $\pi_\theta(a_m|s_m;\theta)$, and a critic neural network with a linear output is used to estimate the value function $V^{\pi_\theta}(s_m)$ using the parameterized $V^{\pi_\theta}(s_m;\theta_{\bm v})$, which is used to evaluate the learned policy $\pi_\theta(a_m|s_m;\theta)$.

The gradient of the expected total cumulative reward with respect to the policy parameters $\theta$ is given by:
\begin{equation}\label{equ:gradient}
		\nabla_\theta \mathbb E_{\pi_\theta}[\sum_{m=0}^{\infty} \gamma^{m}r_m]\!=\!\mathbb E_{\pi_\theta}[\nabla_\theta \log\pi_\theta(a_m|s_m;\theta)A^{\pi_\theta}(s_m, a_m)],
\end{equation}
where $A^{\pi_\theta}(s_m, a_m)=Q^{\pi_\theta}(s_m, a_m)-V^{\pi_\theta}(s_m)$ is the advantage function, which depicts the difference between the expected return when deterministically selecting an action $a_m$ in state $s_m$ and the expected return for actions drawn from policy $\pi_\theta$. It indicates how much better the performance obtained by a specific action than the average level derived by the policy. In practice, the agent samples a trajectory of real experience $(s_m, a_m, r_m, s_{m+1})$ derived by following the policy $\pi_\theta$, and estimates the advantage function $A^{\pi_\theta}(s_m, a_m)$ using the empirically computed $A(s_m, a_m)$. To be specific, we can derive the estimated advantage $A(s_m, a_m)$ by the temporal difference method with $n$-step bootstrapping as:
\begin{equation}\label{equ:advantage}
	A(s_m, a_m;\theta_{\bm v})=\sum_{j=0}^{n-1}\gamma^j r_{m+j} + \gamma^n V(s_{m+n};\theta_{\bm v}) -V(s_{m};\theta_{\bm v}). 
\end{equation}

Correspondingly, we train the critic network parameters $\theta_{\bm v}$ by the following update rule:
\begin{equation}\label{equ:updatecritic}
\theta_{\bm v} \gets \theta_v \!-\! \mu^\prime \sum_m\nabla_{\theta_v}(A(s_m, a_m;\theta_{\bm v}))^2,
\end{equation}
where $\mu^\prime$ is the learning rate of the critic network. Then, we can update the policy parameters $\theta$ of the actor network by
\begin{equation}\label{equ:updateactor}
\theta \gets \theta + \mu \sum_m\nabla_\theta\log\pi_\theta(a_m|s_m;\theta)A(s_m, a_m)+\varphi \nabla_\theta h(\pi_\theta(\cdot|s_m))
\end{equation}
with $\mu$ being the learning rate of the actor network. Intuitively, the direction $\nabla_\theta\log\pi_\theta(a_m|s_m;\theta)$ prescribes how to adjust the parameters $\theta$ to increase $\pi_\theta(a_m|s_m;\theta)$, namely, the probability of performing action $a_m$ at the state $s_m$, along with a step size $A(s_m, a_m)$ reflecting the advantage value for taking action $a_m$ in state $s_m$, which guides the network to reinforce actions that lead to better returns. Here, $h(\cdot)$ is the entropy of the policy, which is used to encourage exploration (give preference to try different actions to discover good policies) by pushing $\theta$ in the direction of higher entropy. The hyperparameter $\varphi$ controls the strength of exploration that is set to a large value at the beginning of training and decreases over time. The pseudocode for the A3C algorithm can be found in~\cite{A3C}.

\section{Experiments}
In this section, we evaluate the performance of the proposed two-level decision making framework for multiuser adaptive streaming in heterogeneous networks. Specially, we take the weighted sum mean-square error minimization (WMMSE) algorithm~\cite{WMMSE} as the baseline scheme in wireless resource allocation, which devotes to maximizing the sum of transmission rate of each user at each time slot without considering video quality or long-term performance. Meanwhile, for rate adaptation part, we compare the proposed deep reinforcement learning (DRL)-based algorithm that maximizes the long-term user-perceived Quality of Experience (QoE) with two widely used schemes, namely, rate-based (RB)~\cite{rate-based} and buffer-based (BB)~\cite{buffer-based} rate adaptation mechanisms. The former predicts the future throughput using the harmonic mean of the historically measured throughput trace, and then selects the highest available bitrate which is below the predicted throughput. The other one chooses the chunk bitrate with the goal for keeping the buffer occupancy above a present level. As a result, rebuffering event will never happen in this case at the cost of lower video quality in some cases. In a nutshell, we illustrate the effectiveness of our proposed cross-layer combination of QDDRA algorithm for wireless resource allocation and DRL-based rate adaptation logic (called QDDRA$\_$DRL) over the following five schemes: QDDRA$\_$BB, QDDRA$\_$RB, WMMSE$\_$DRL, WMMSE$\_$BB and  WMMSE$\_$RB, which combine different methods in physical and application layer respectively. 
%\IEEEpubidadjcol

\subsection{Simulation Settings}
\begin{table}[t]
\caption{Configuration of Parameters in Radio Model.}\footnotesize
\centering
\begin{tabular}{|c|c|c|}
 	\hline 
	\textbf{Parameter}&\textbf{Description}&\textbf{Value}\\
	\hline
	$T_{slot}$ & Length of a time slot in seconds & 0.04\\
	\hline
	$D$ & Cell radius in meters & 100 \\
    \hline 
    $D_{min}$ & Minimum allowable radius in meters & 10 \\
    \hline
	$P_k$ & Power budget of base station $k$ in watts  & 4\\
	\hline
	$B$ & Channel bandwidth in Hz  & $10^6$\\
	\hline
	$\sigma^2$ & Background noise in watts & 1\\
	\hline
	$\Gamma$ & SNR gap & 1.34\\
	\hline
    $f_d$ & Maximum Doppler frequency in Hz & 10\\
    \hline
    $1-\beta$ & Cumulative video quality weight & 0.9 \\
    \hline
\end{tabular}
\label{tab:radioparameter}
\end{table}

\begin{table}[t]
\caption{Parameters Setting in Rate Adaptation.}\footnotesize
\centering
\begin{tabular}{|c|c|c|}
 	\hline 
	\textbf{Parameter}&\textbf{Description}&\textbf{Value}\\
	\hline
	$T_{chunk}$ & Chunk length in seconds  & 2\\
	\hline
	$b_{max}$ & Buffer size in seconds  & 30\\
	\hline
	 $\lambda$ & Quality variation penalty weight & 0.5\\
	 \hline
	$\rho$ & Rebuffering penalty weight & 4 \\
    \hline 
    $\mu$ & Learning rate of actor network & $10^{-5}$ \\
    \hline
    $\mu^\prime$ & Learning rate of critic network & $10^{-4}$ \\
    \hline  
    $\gamma$ & Discount factor & 0.99\\
    \hline 
    $\varphi$ & Entropy factor & 0.5\\
    \hline
\end{tabular}
\label{tab:qoeparameter}
\end{table}

We employ a block fading model to describe the downlink channel gain $\mathbf H_{i_k, k}^t$ between base station $k\in\mathcal K$ and mobile user $i_k\in\mathcal I$, which is composed of large-scale fading component and small-scale Raleigh fading component. We adopt the log-distance path loss model with log-normal shadowing to represent large-scale fading, which is affected by the distance between the base station and the user, i.e., user mobility. To this end, we stipulate that users will change their positions every few seconds (set to 5 seconds in our experiment) in a random direction and reasonable walking distance (relative to normal walking speed). Besides, we define a minimum allowable radius $D_{min}$ to prevent the transmitter-receiver distance from being too close, that is, there is no users located in the inner region of a base station with a radius of $D_{min}$. As to the small-scale Raleigh fading, we use the Jakes' model~\cite{Jakes} to depict its variation expressed as a first-order complex Gauss-Markov process, i.e., 
\begin{equation}\label{small-scale Raleigh fading}
\mathbf G^{t}_{i_k, k}=\zeta \mathbf G^{t-1}_{i_k, k}+\bm \xi^{t}_{i_k, k},\quad \forall k\in\mathcal K, i_k\in\mathcal I.
\end{equation}
The channel innovation process $\bm \xi^{1}_{i_k, k}, \bm \xi^{2}_{i_k, k},\dots$ are composed of some independent and identically distributed circularly symmetric complex Gaussian (CSCG) random variables with distribution $\mathcal {CN}(0, 1-\zeta^2)$, and the correlation factor $\zeta=J_0(2\pi f_dT_{slot})$, where $J_0(\cdot)$ and $f_d$ denote the zero-order Bessel function of the first kind and the maximum Doppler frequency respectively. The initial Raleigh fading coefficients are also CSCG variables with unit variance. The main parameters of radio model are listed in Table \ref{tab:radioparameter}.

We use H.264/MPEG-4 codec at the DASH server to encode three test video sequences~\cite{Video} ($F=3$, BigBuckBunny, SitaSingstheBlues and DucksTakeOff) with 1080p resolution (1920$\times$1080) and 30 fps (frame per second) frame rate. All of them are encoded into $L=6$ different representations with encoding bitrate set as $\{0.3, 0.75, 1.2, 1.85, 2.85, 3.2\}$ Mbps\footnote{Many server-side representation selection algorithms like ~\cite{lichenglin}, \cite{Laura2} can optimally decide the number of representations and the corresponding bitrate set by appropriately choosing the settings (e.g., encoding parameters, frame rate, resolution, etc.) such that the total system utility (e.g., aggregate users' satisfaction) is maximized. For simplicity, we assume that all video chunks have the same representations.}. Besides, these videos are further divided into small chunks, each of which contains 2 seconds of video contents. We adopt PSNR as the video quality metric and measure the quality of each chunk by computing the average PSNR value of frames within the chunk. In QoE evaluation, we carefully set the rebuffering penalty weight such that the decrease of QoE induced by one second rebuffering time is the same as reducing video quality by 4 dB. As illustrated in Fig. \ref{fig:A3C}, the inputs of the actor and critic neural network consist of the experienced throughput and the download time for the past $n=8$ chunks, the complexity of the chunks to be downloaded that is signified by its PSNR values of $L$ available bitrates, etc. Furthermore, the two networks have the same architecture except that the actor network has a softmax output while a linear output for the critic network. The specific parameters used to train the two neural networks are presented in Table \ref{tab:qoeparameter}.

\subsection{Results}

\begin{figure*}[!h]
\centering
\subfloat[Overall average transmission rate of all users.]{\includegraphics[width=0.48\linewidth]{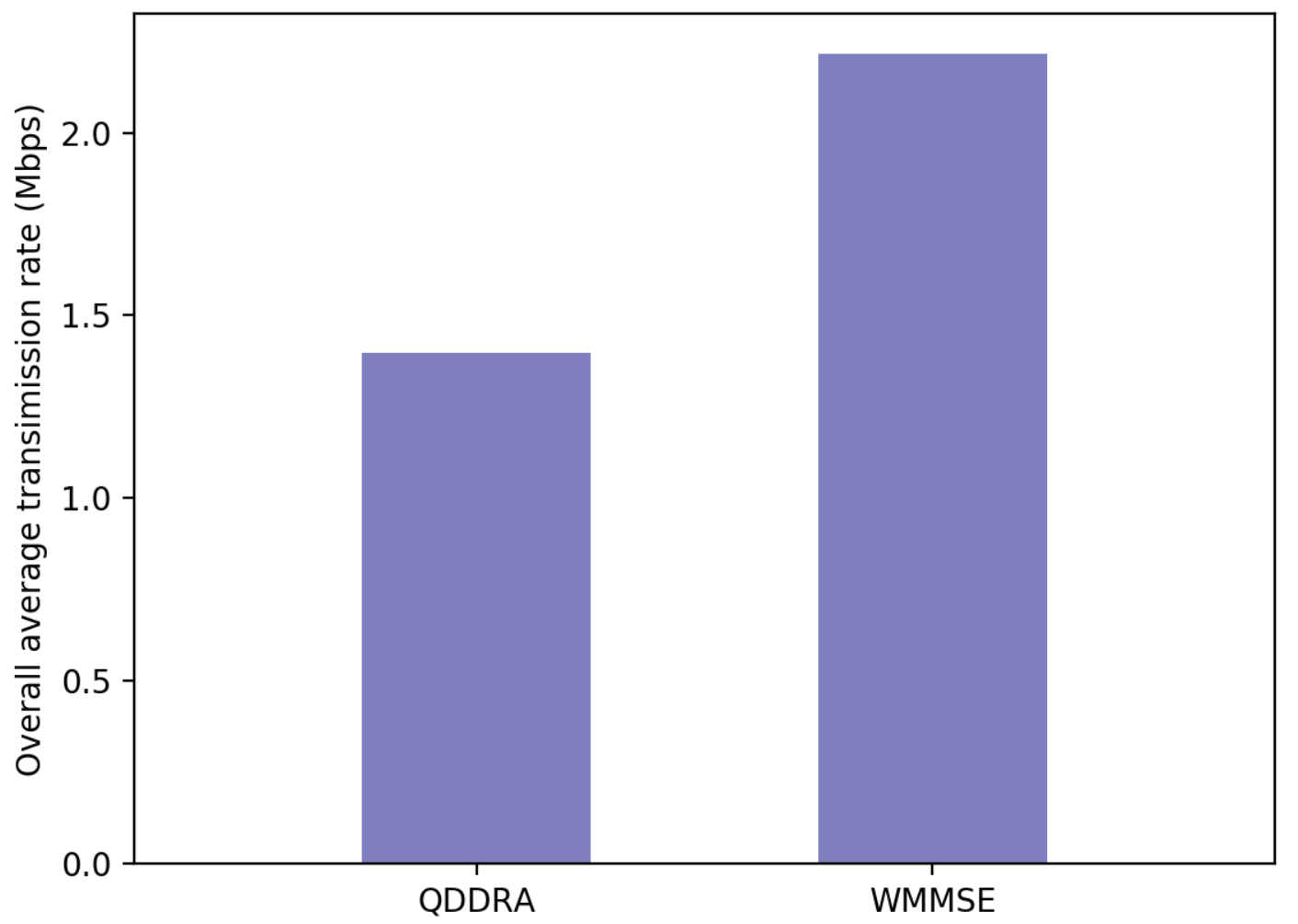}
\label{fig:transmission rate a}}
\hfill
\subfloat[User unfairness in terms of allocated transmission rate.]{\includegraphics[width=0.48\linewidth]{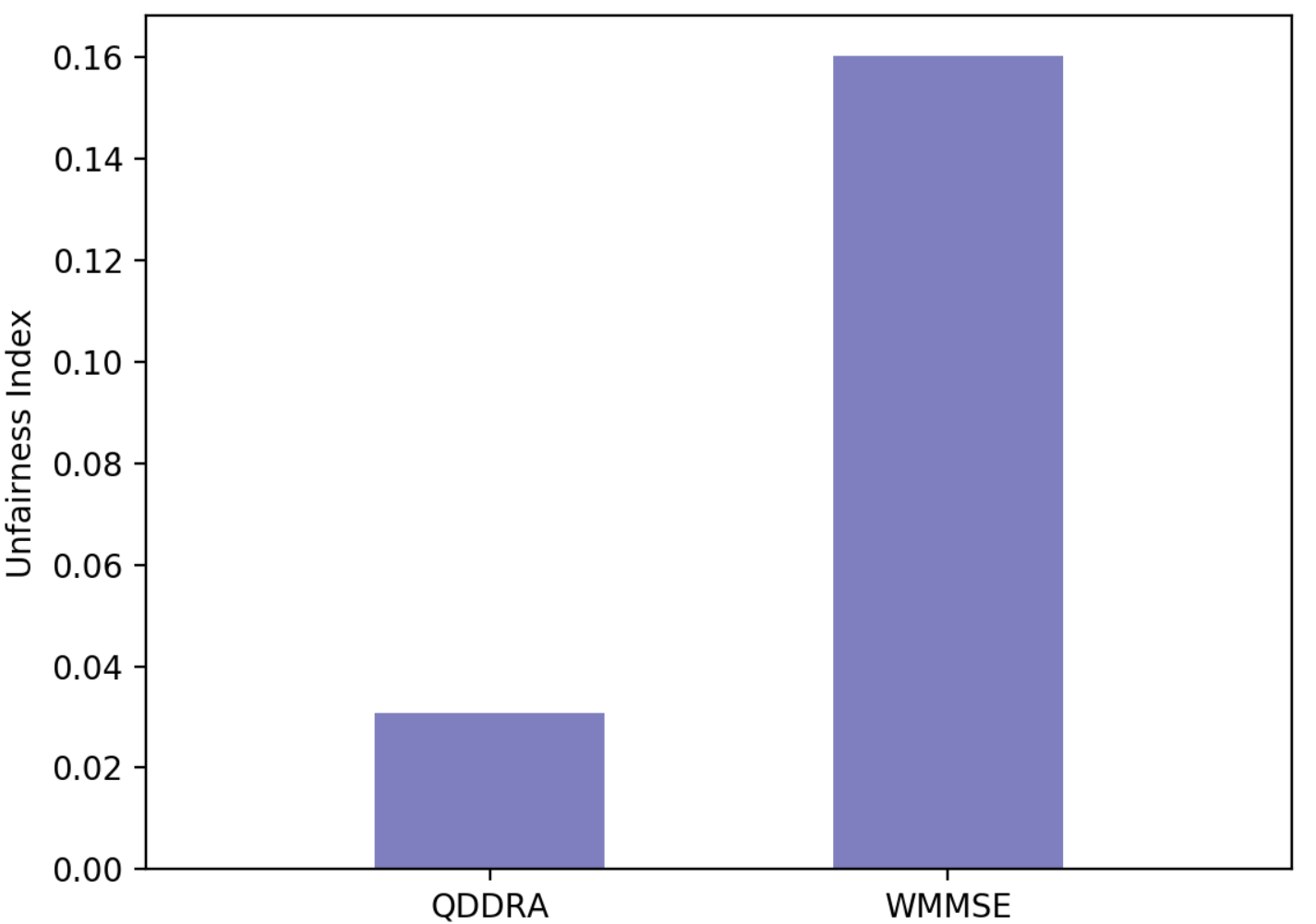}
\label{fig:transmission rate b}}
\hfill
\subfloat[Average transmission rate per user using QDDRA.]{\includegraphics[width=0.48\linewidth]{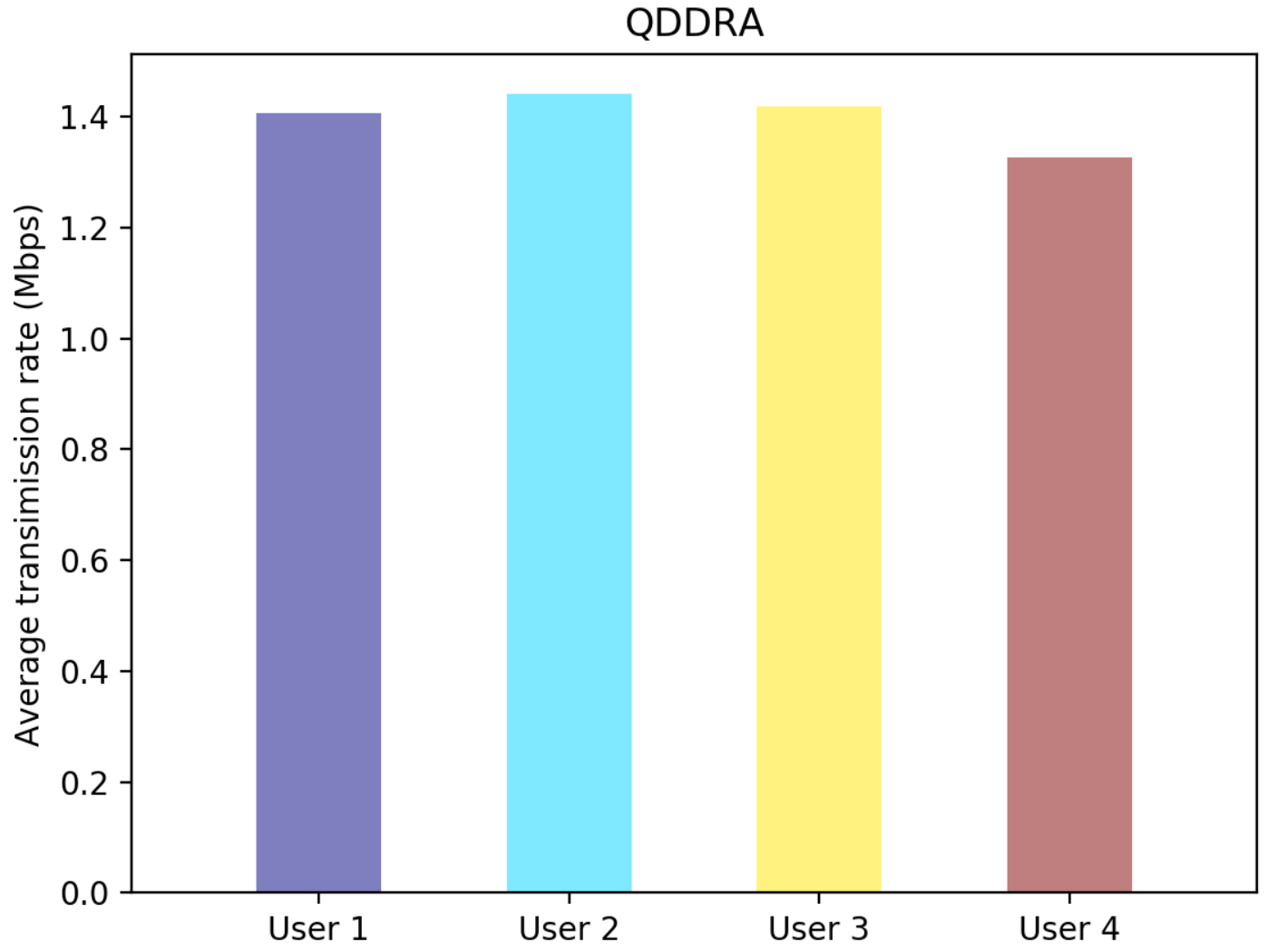}
\label{fig:transmission rate c}}
\hfill
\subfloat[Average transmission rate per user using WMMSE.]{\includegraphics[width=0.48\linewidth]{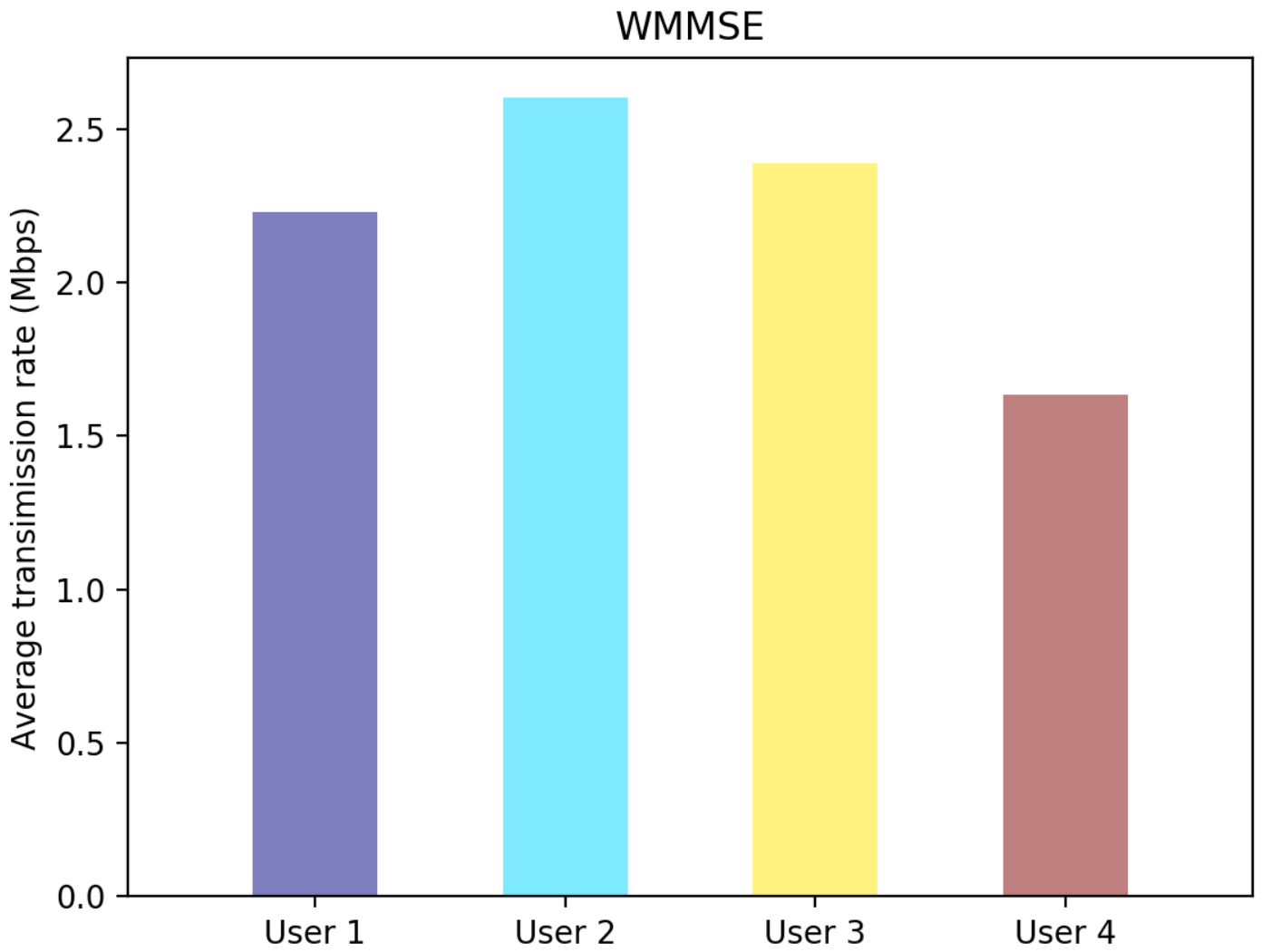}
\label{fig:transmission rate d}}
\caption{Comparison of allocated transmission rate achieved by QDDRA and WMMSE algorithms.}
\label{fig:transmission rate}
\end{figure*}

\begin{figure*}[!h]
\centering
\subfloat[Overall average QoE of all users.]{\includegraphics[width=0.48\linewidth]{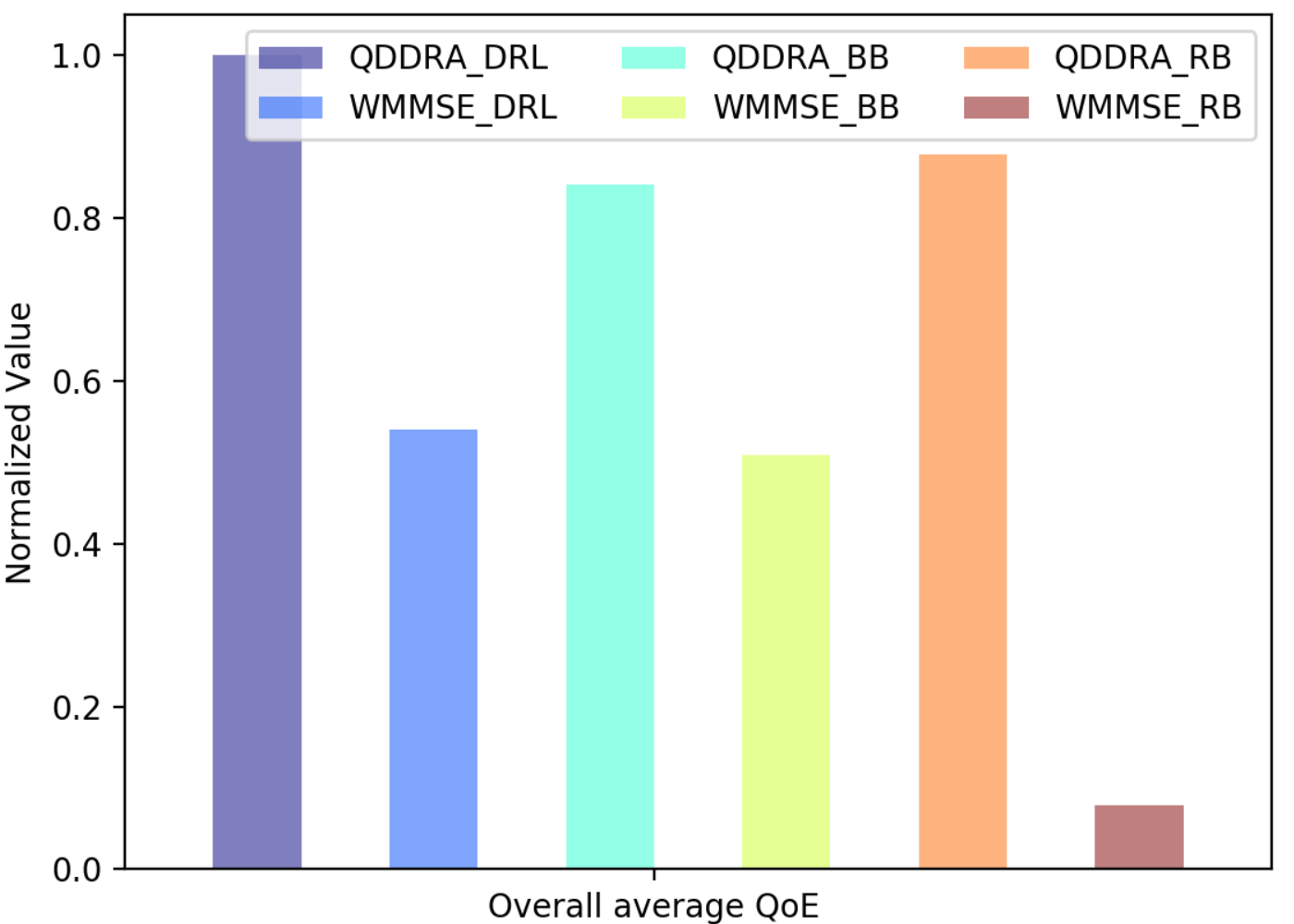}
\label{fig:overall performance a}}
\hfill
\subfloat[The CDF of overall average QoE of all users.]{\includegraphics[width=0.48\linewidth]{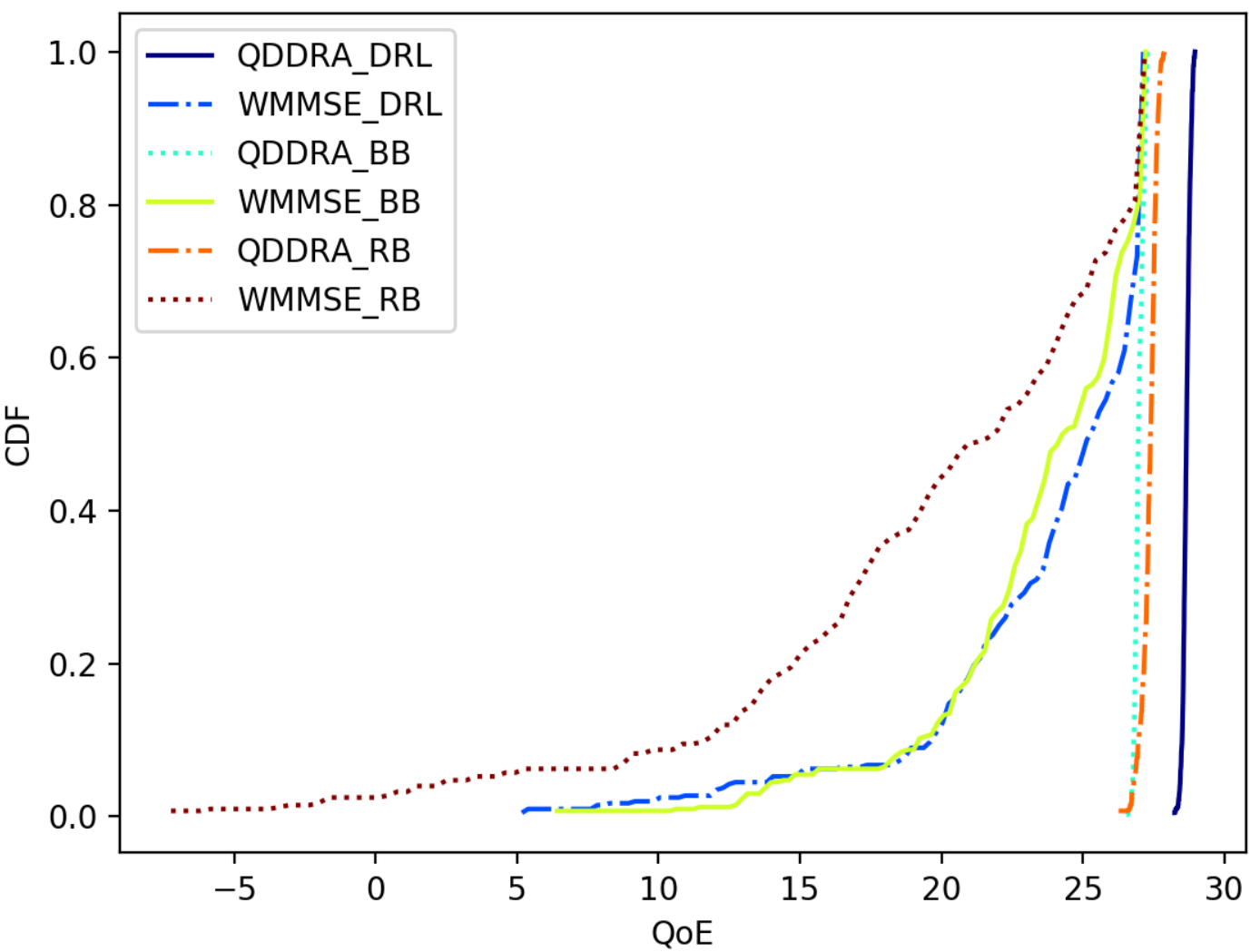}
\label{fig:overall performance b}}
\hfill
\subfloat[Overall average performance on the individual components.]{\includegraphics[width=0.48\linewidth]{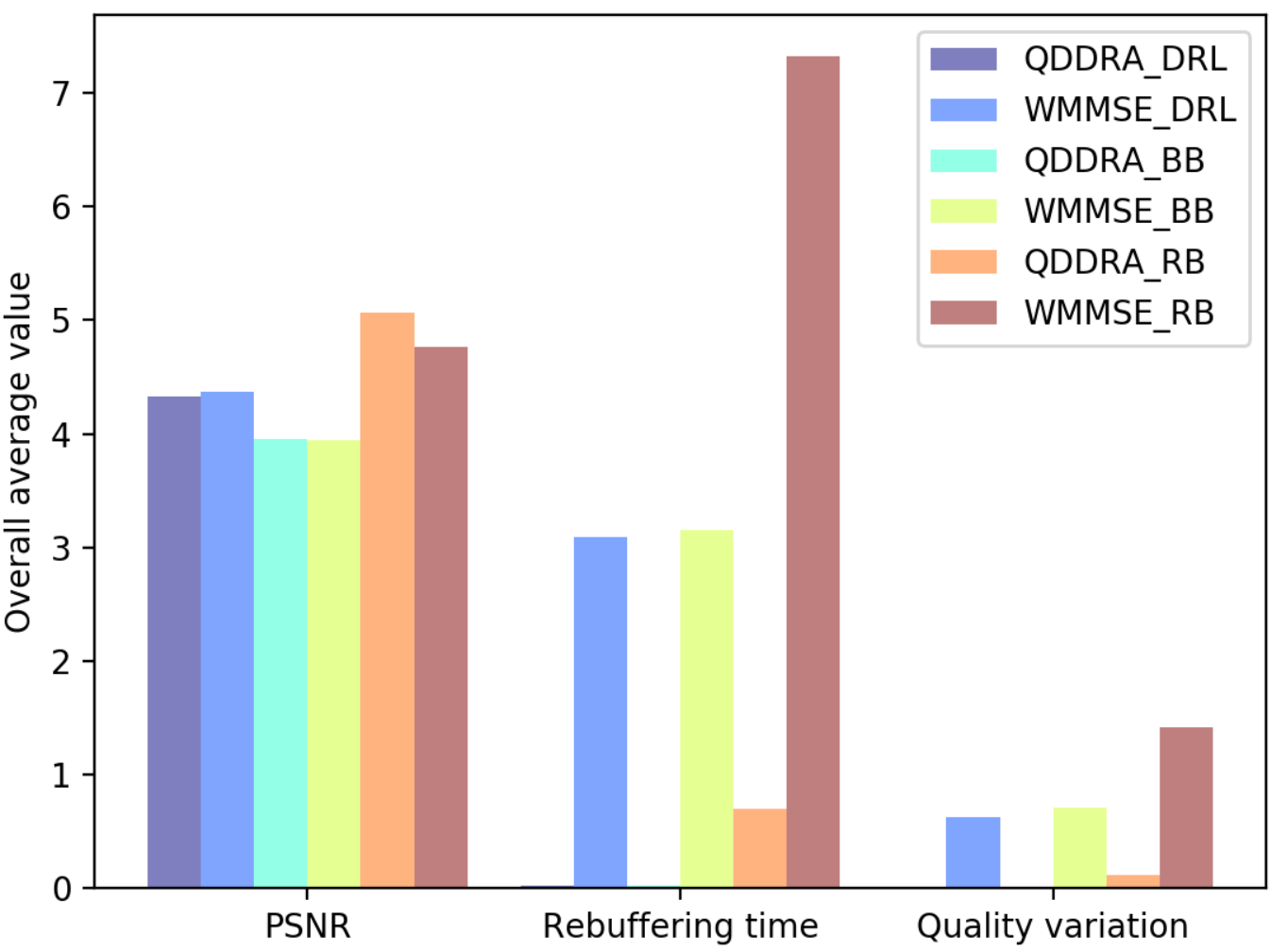}
\label{fig:overall performance c}}
\hfill
\subfloat[Unfairness in terms of user-perceived average QOE.]{\includegraphics[width=0.48\linewidth]{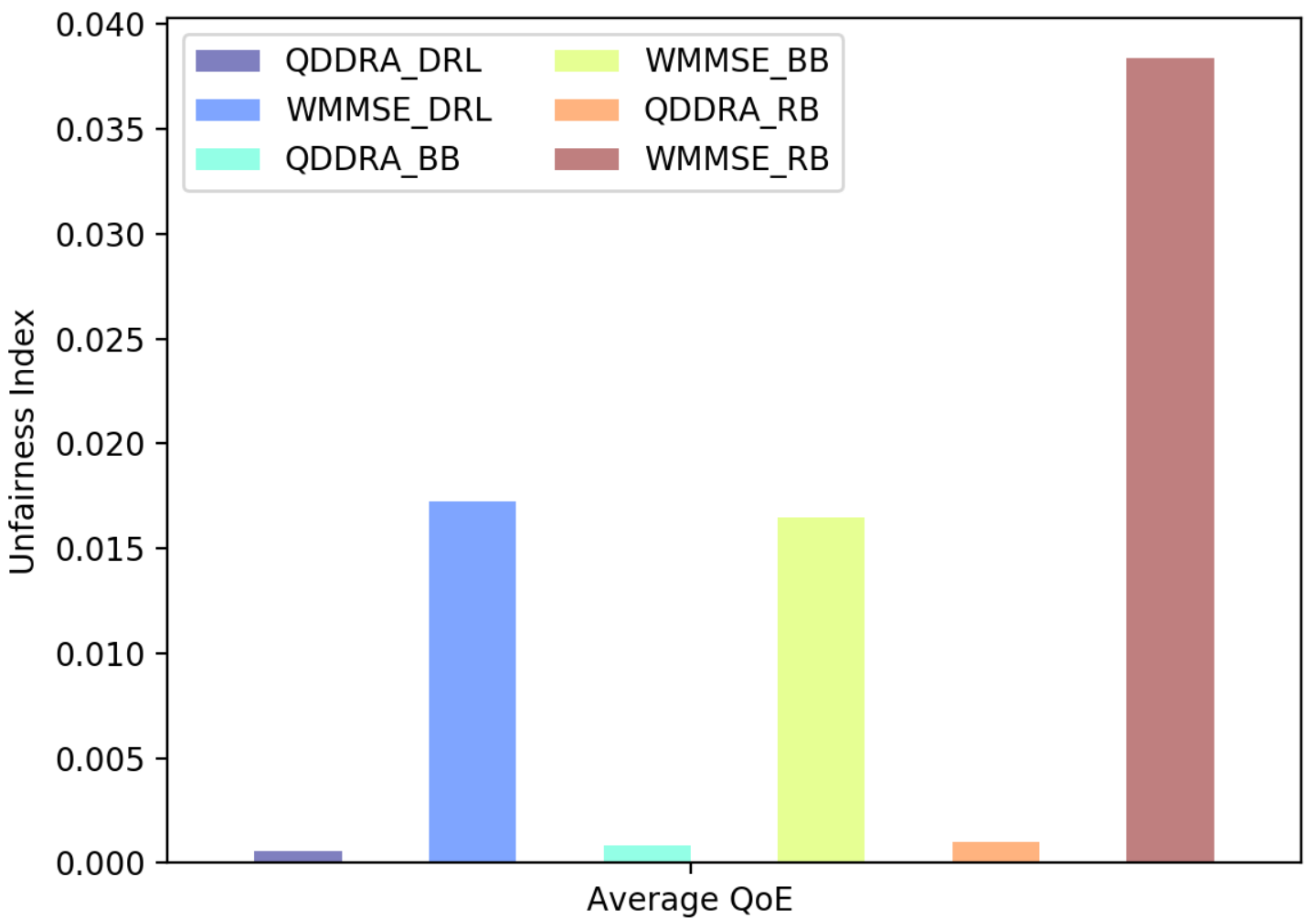}
\label{fig:overall performance d}}
\caption{Comparison of overall average performance achieved by 6 different algorithms.}
\label{fig:overall performance}
\end{figure*}

\begin{figure*}[!h]
\centering
\subfloat[Average QoE per user]{\includegraphics[width=0.48\linewidth]{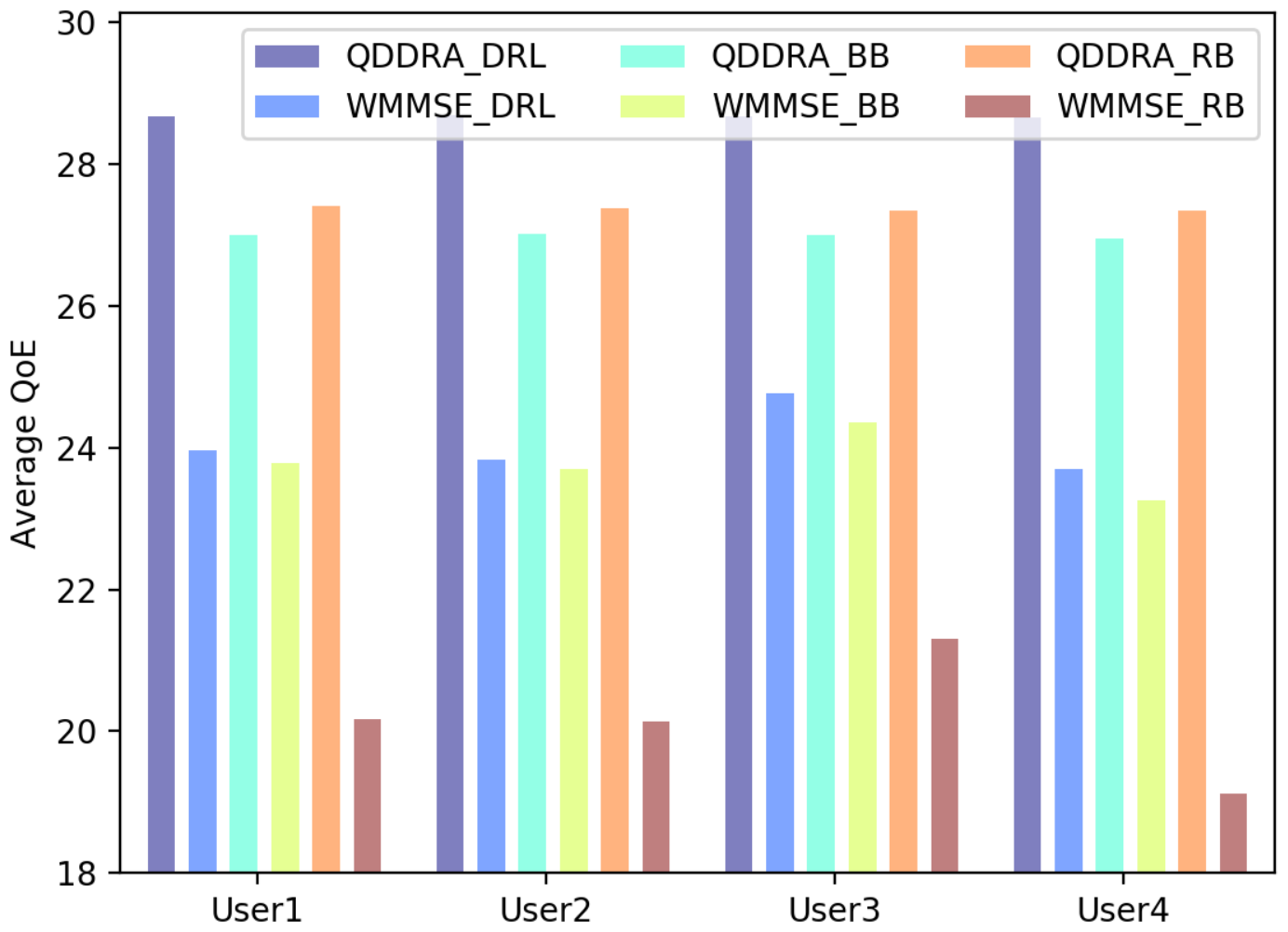}
\label{fig:per user performance a}}
\hfill
\subfloat[Average PSNR per user]{\includegraphics[width=0.48\linewidth]{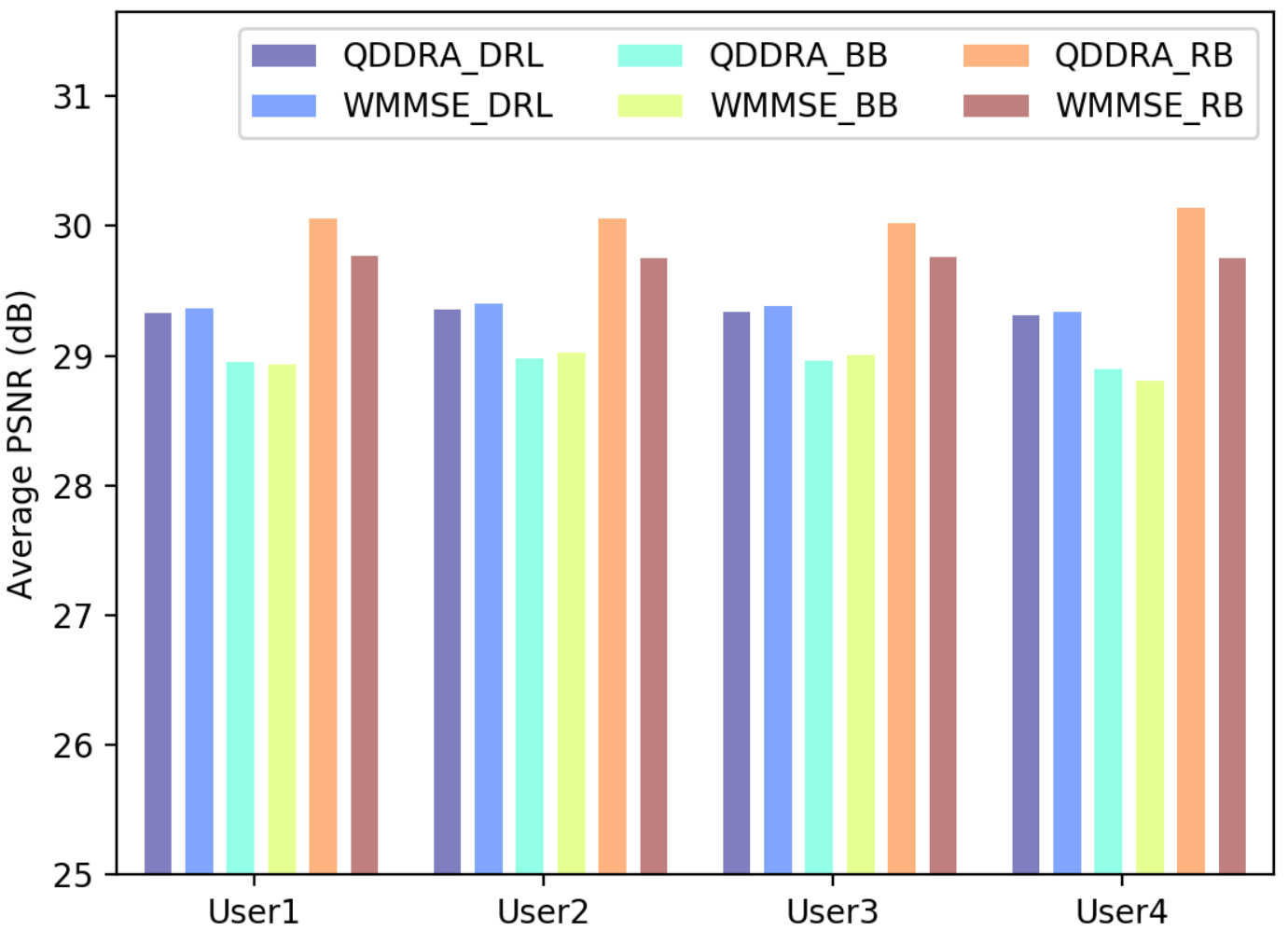}
\label{fig:per user performance b}}
\hfill
\subfloat[Average quality variation per user]{\includegraphics[width=0.48\linewidth]{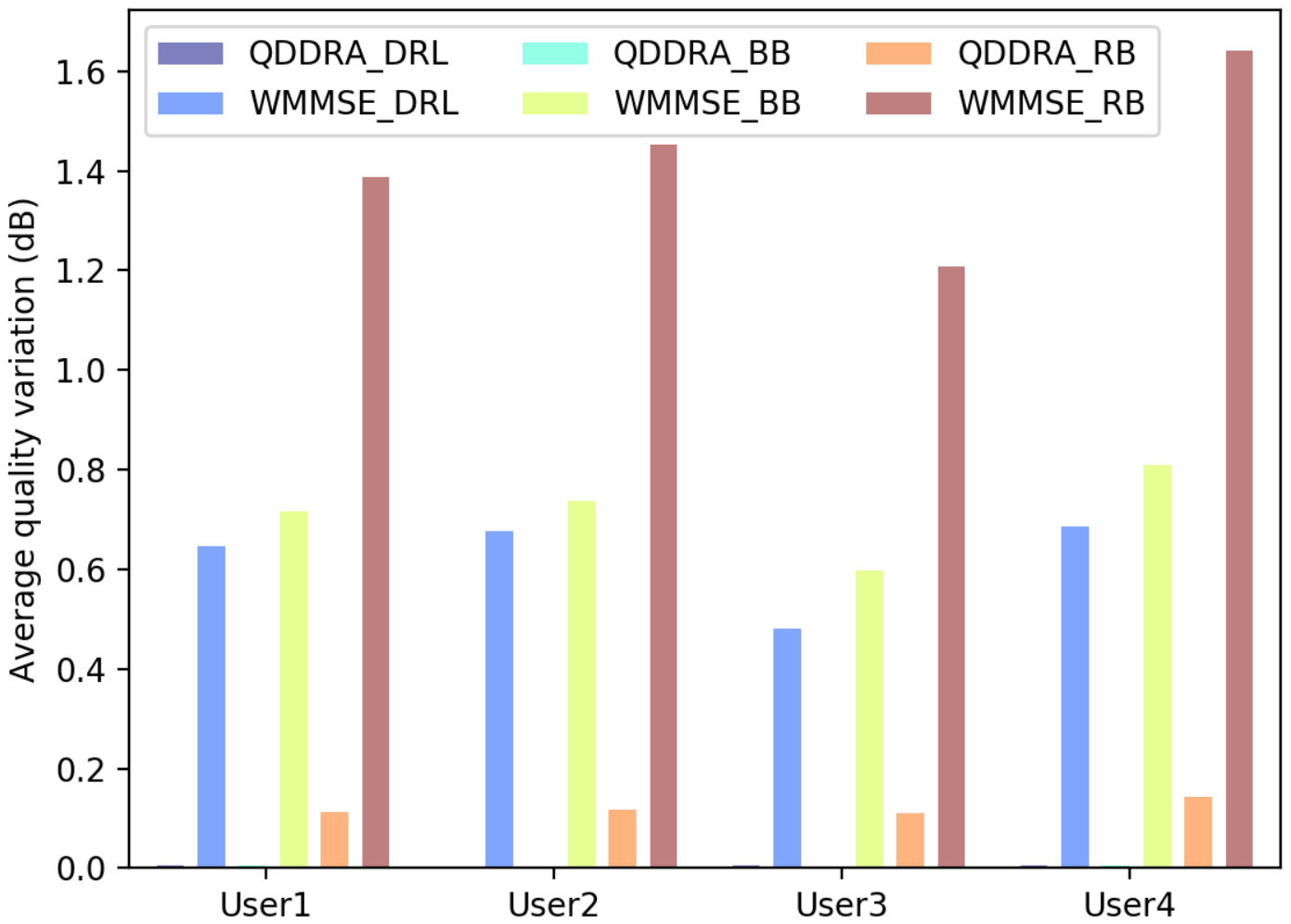}
\label{fig:per user performance c}}
\hfill
\subfloat[Average rebuffering time per user]{\includegraphics[width=0.48\linewidth]{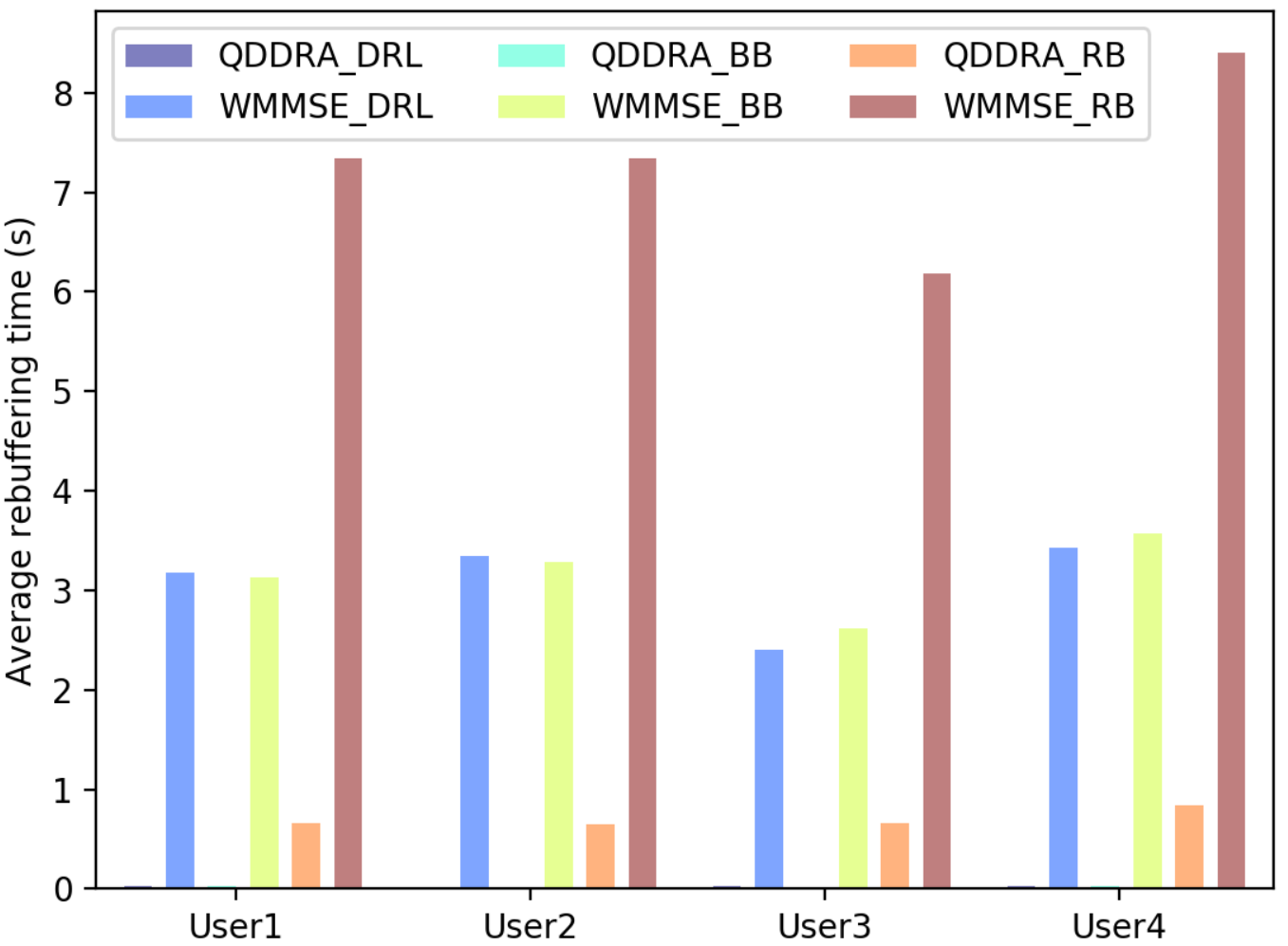}
\label{fig:per user performance d}}
\caption{Comparison of average performance per user achieved by 6 different algorithms.}
\label{fig:per user performance}
\end{figure*}

\begin{figure*}[!h]
\centering
\subfloat[User 1]{\includegraphics[width=0.48\linewidth]{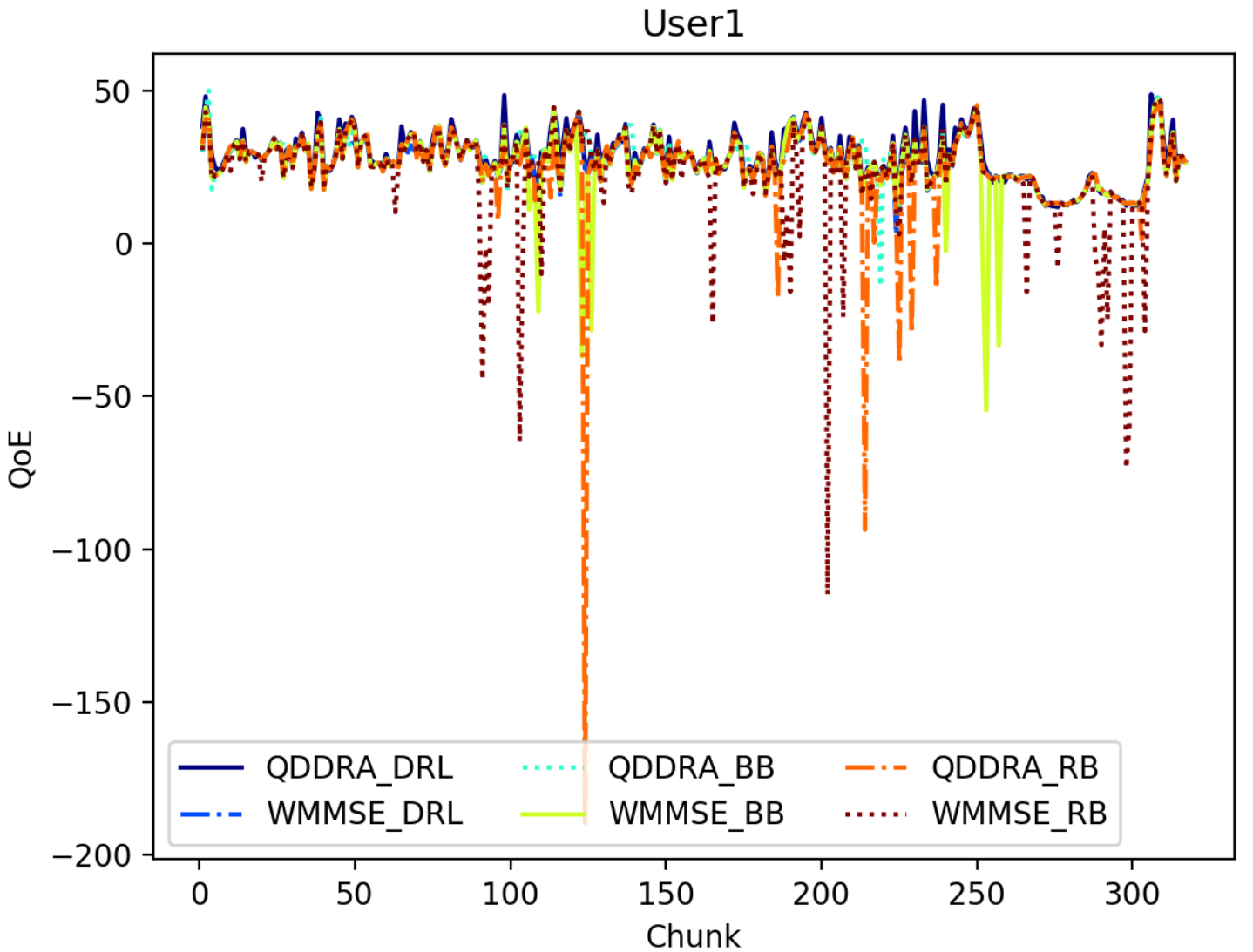}
\label{fig:chunk-by-chunk QoE a}}
\hfill
\subfloat[User 2]{\includegraphics[width=0.48\linewidth]{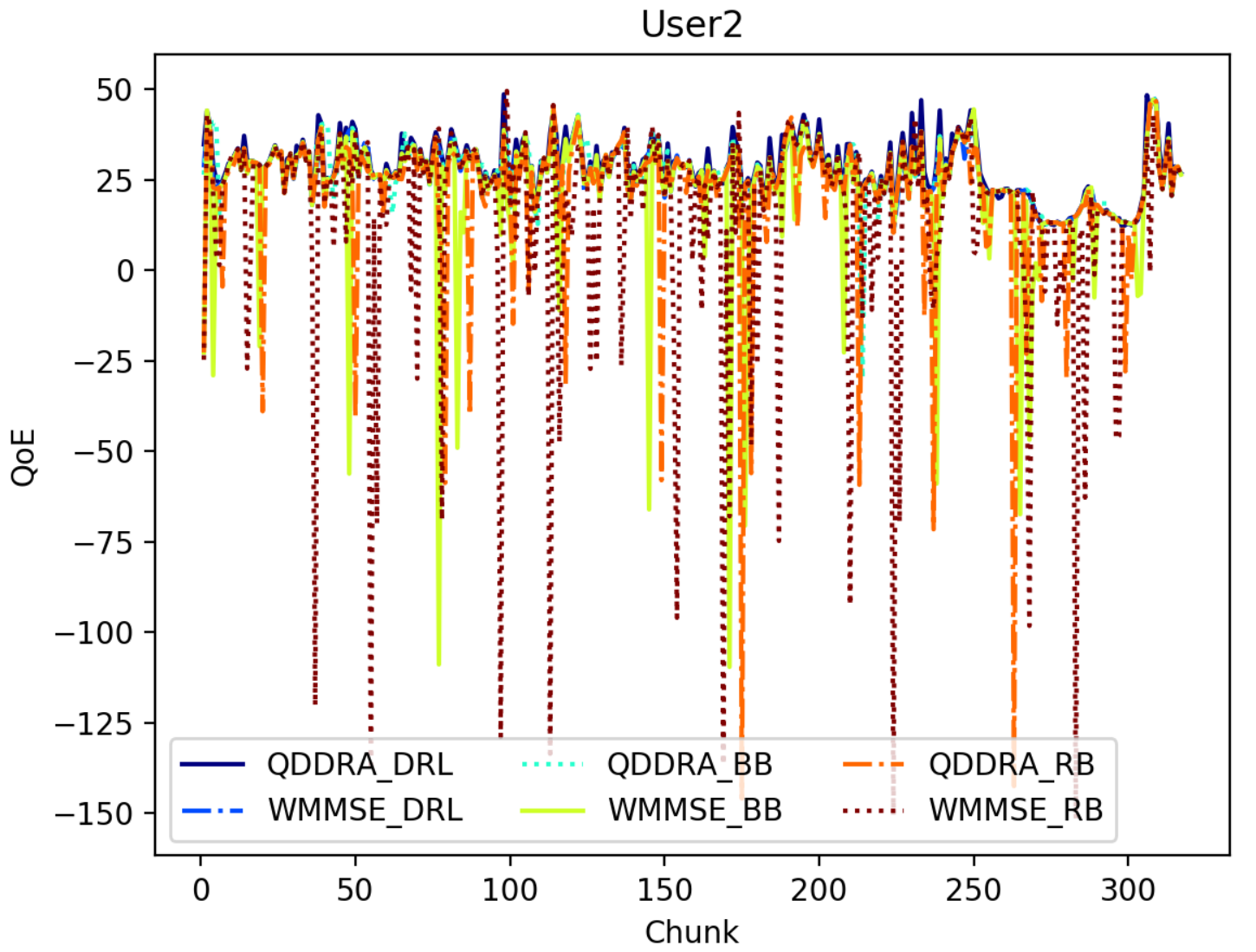}
\label{fig:chunk-by-chunk QoE b}}
\hfill
\subfloat[User 3]{\includegraphics[width=0.48\linewidth]{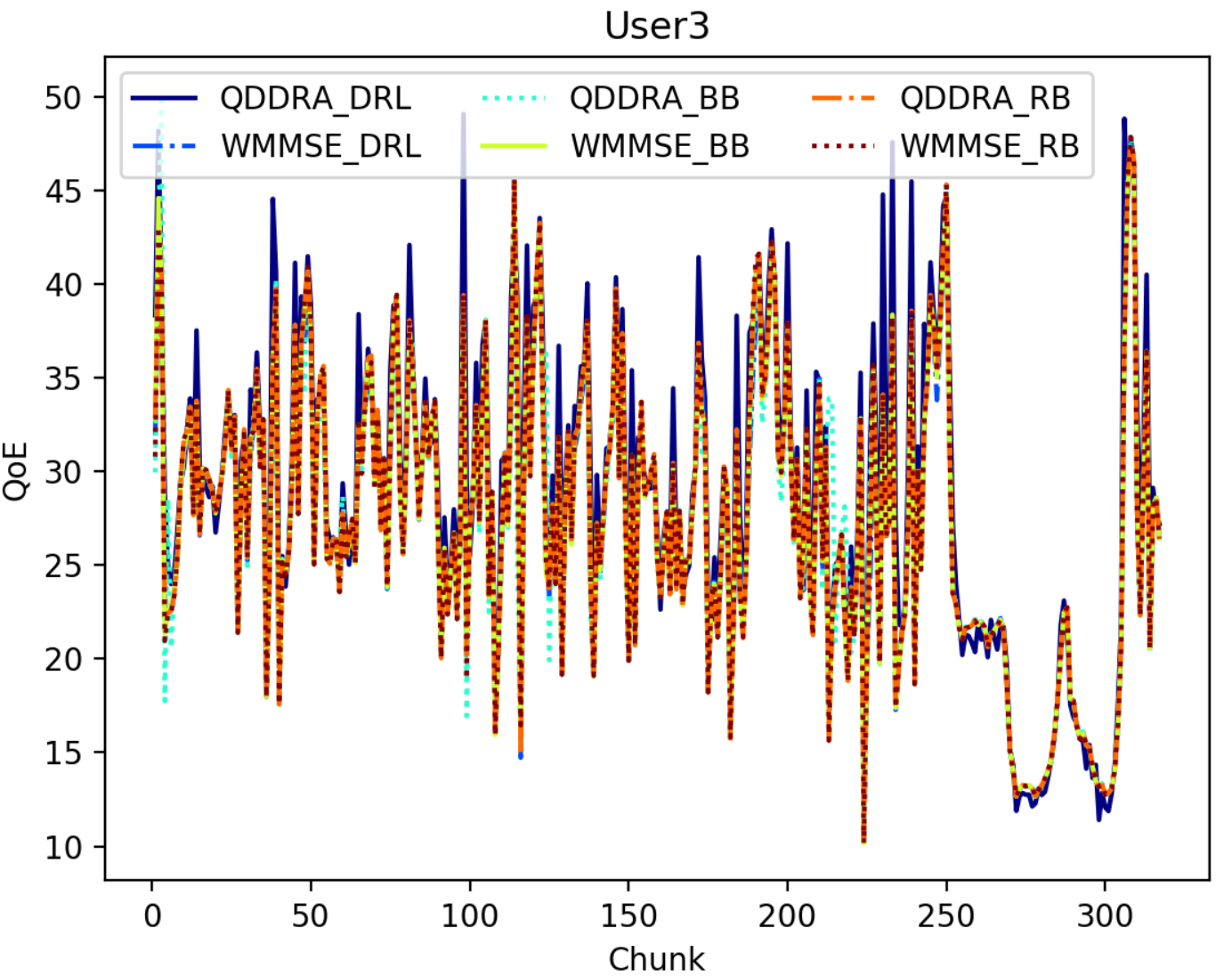}
\label{fig:chunk-by-chunk QoE c}}
\hfill
\subfloat[User 4]{\includegraphics[width=0.48\linewidth]{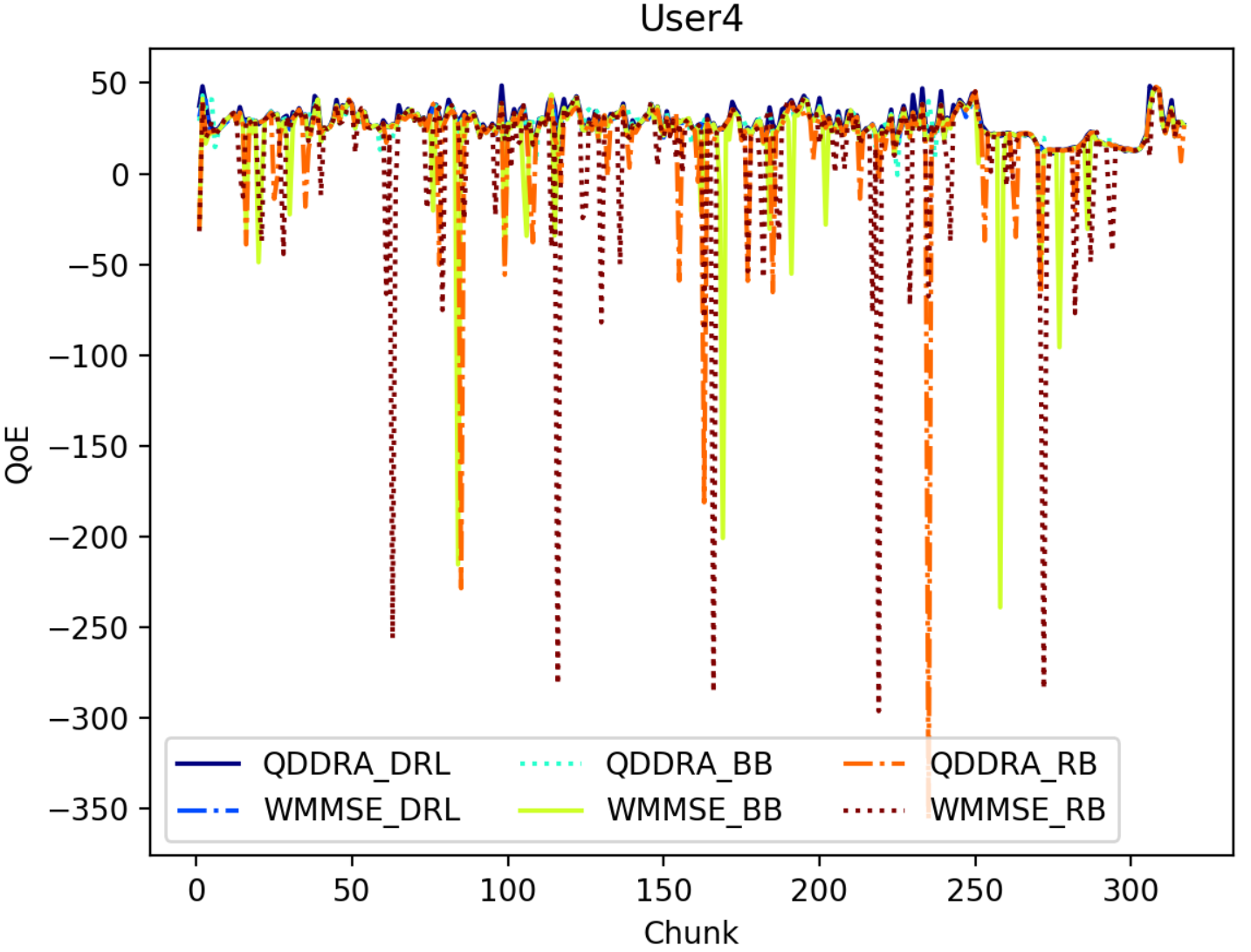}
\label{fig:chunk-by-chunk QoE d}}
\caption{Comparison of the chunk-by-chunk QoE per user achieved by 6 different algorithms.}
\label{fig:chunk-by-chunk QoE}
\end{figure*}

In this subsection, we compare the performance of the proposed QDDRA$\_$DRL with the other five baselines. First, we consider a single cell scenario where a base station equipped with a single antenna serves four single-antenna users with single-input single-out (SISO) interference channel. Moreover, we assume that all four users request playback for the same video (BigBuckBunny).  We carry out the QDDRA and WMMSE algorithms in Matlab to generate some network throughput traces, each of which logs the allocated transmission rate for each user over 300 seconds at a 1 second granularity. To be specific, we randomly assign an initial position for each user in the beginning of each trace, and then utilize the two resource allocation algorithms to compute the transmission rate per user over time. The channel gain changes following the model described in last subsection. We generate 100 traces for training and and another 100 for testing. Then, we exploit the throughput traces in training set to train the A3C-based rate adaptation logic using TensorFlow, and test the user-perceived QoE when playing the video with network traces in the testing set. Moreover, we report the average QoE of all chunks, that is, the total QoE divided by the number of chunks within the video, which is computed when the users watch videos over 100 test traces.

Fig. \ref{fig:transmission rate} compares the allocated transmission rate derived by QDDRA and WMMSE algorithms. Although the WMMSE algorithm achieves higher overall average transmission rate as shown in Fig. \ref{fig:transmission rate a}, the user unfairness in terms of allocated transmission rate is dramatically greater than the proposed QDDRA algorithm, which is illustrated in Fig. \ref{fig:transmission rate b}. Here, similar to \cite{FESTIVE}, we measure the user unfairness using $\sqrt{1-JainFair}$ where $JainFair$ is the Jain fairness index~\cite{JainFair} of allocated average transmission rate over four users. 

Figs. \ref{fig:transmission rate c} and \ref{fig:transmission rate d} show the average transmission rate per user based on QDDRA and WMMSE, respectively, which suggest the achieved performance and their individual optimization objectives are consistent: WMMSE aims to maximize the sum of transmission rate for all users at each time slot, and QDDRA aims at maximizing the accumulated average video quality of each user over a period of time under proportional fairness consideration. As shown later, the user-perceived QoE is not aligned with overall transmission rate. The main reason is that WMMSE may assign a user with excessively high transmission rate but at the same time allocate a very low (even zero) rate to the other users for the sake of sum-rate maximization. This would be likely to cause rebuffering events for users with low transmission rate. At the same time, the user allocated with more wireless resources may not benefit from it due to the limitation on buffer size.  

Fig. \ref{fig:overall performance a} shows the overall average normalized QoE of all users, and Fig. \ref{fig:overall performance b} presents the corresponding cumulative distribution function (CDF). It can be seen that the performance of our proposed QDDRA$\_$DRL exceeds the remaining methods with respect to the overall average QoE. In addition, QDDRA-based methods have better performance than the WMMSE-based methods---showing the effectiveness of our video streaming-tailored resource allocation algorithm. We further analyze the overall average performance on the individual components defined in Eq. \eqref{equ:QOE}, as illustrated in Fig. \ref{fig:overall performance c}. Since the PSNR has a much bigger value than others, here, we reduce PSNR value by 25 dB for all users to better present the differences achieved by 6 methods with respect to the three items. The QDDRA$\_$DRL performs well in all three items, which means a good tradeoff on the video quality, quality variation as well as rebuffering time is achieved. %QDDRA$\_$BB has no rebuffering events or quality fluctuation at the cost of having the lowest video quality. WMMSE$\_$RB has the worst performance among all methods for failing to precisely predict the future throughput in rapidly changing network environment. 
%However, QDDRA$\_$RB is much improved relative to WMMSE$\_$RB.
%In comparison, WMMSE$\_$DRL can better learn the network variation by interacting with the environment, and adjust the action in the selected representation. 
All WMMSE-based methods cannot achieve balanced performance over the three components, leading to less appealing QoEs. Fig. \ref{fig:overall performance d} further compares the unfairness in the user-perceived average QoE, where one can see that the QDDRA-based methods exhibits much better fairness.

The average performance per user achieved by the six different algorithms is illustrated in Fig. \ref{fig:per user performance}. It is observed in Figs. \ref{fig:per user performance a} that QDDRA$\_$DRL performs well in all cases in terms of average QoE. %On one hand, due to fairness consideration, QDDRA algorithm provides all users with a basic transmission rate to support their video delivery. On the other hand, DRL-based can learn how to maximize the selected video quality, and at the same time, can also learn how much buffer is necessary to alleviate the risk of rebuffering event based on the changing network conditions. 
The average PSNR, quality variation as well as rebuffering time of each user are shown in Figs. \ref{fig:per user performance b}, \ref{fig:per user performance c}, \ref{fig:per user performance d} respectively. In order to observe the QoE changes between chunks over time, we show in Fig. \ref{fig:chunk-by-chunk QoE} the chunk-by-chunk QoE of each user when watching the video over a randomly picked trace.

Further, we consider a more complex multiple-input multiple-out (MIMO) interference channel scenario composed of four cells, where each base station equipped 3 transmit antennas serves three users, each of which has 2 receive antennas. In addition, the users in the same cell watch the different videos. The average performance comparison with respect to QoE, PSNR, quality variation and buffering time is summarized in Table \ref{tab:multicell}. The intra-cell fairness (evaluated by unfairness index in the perceived average QoE of users within the same cell) as well as total fairness (evaluated by unfairness index in the perceived average QoE of all users) are listed in Table \ref{tab:multicellunfairness}. Similar to the single cell scenario, our proposed QDDRA$\_$DRL exhibits the best performance in terms of users' QoEs and fairness.

\begin{table*}[htbp]
\caption{Comparison of Performance Achieved by Different Algorithms in Multicell Scenario.}\footnotesize
\centering
\begin{tabular}{|c|c|c|c|c|c|}
\hline 
  \textbf{Cell} & \textbf{Algorithm} & \textbf{Average QoE} &  \textbf{Average PSNR} &  \textbf{Average quality variation} & \textbf{Average buffering time}   \\
 \hline
\multirow{6}{*}{1} & \emph{\textbf{QDDRA$\_$DRL}} & \emph{\textbf{33.02}} &  \emph{\textbf{34.54}} & \emph{\textbf{0.16}} & \emph{\textbf{0.95}}      \\
\cline{2-6}
                  & WMMSE$\_$DRL &32.77 &34.68  &0.32  &1.58           \\
                  \cline{2-6}
                  & QDDRA$\_$BB & 31.85 &33.22  &0.02  &0.08                 \\
                  \cline{2-6}
                  & WMMSE$\_$BB & 31.51 & 33.36 &0.14  & 0.55          \\
                  \cline{2-6}
                  & QDDRA$\_$RB & 31.79 & 34.59 &0.30  &1.43             \\
                  \cline{2-6}
                  & WMMSE$\_$RB & 31.37 &35.18  & 0.51 &2.41            \\
\cline{1-6}
\multirow{6}{*}{2} &  \emph{\textbf{QDDRA$\_$DRL}} &  \emph{\textbf{33.48}} & \emph{\textbf{34.66}}  &  \emph{\textbf{0.11}} & \emph{\textbf{0.61}}   \\
    \cline{2-6}
                  & WMMSE$\_$DRL & 33.81 &34.88  &0.16  &0.74           \\
      \cline{2-6}
                  & QDDRA$\_$BB & 32.16 &33.54  &0.03  &0.12                  \\
         \cline{2-6}
                  & WMMSE$\_$BB & 32.30  &33.78  &0.04  &0.19                  \\
          \cline{2-6}
                  & QDDRA$\_$RB & 32.51 &34.90  &0.22  &1.02             \\
           \cline{2-6}
                  & WMMSE$\_$RB & 32.81  &35.49  &  0.25&1.24               \\
\cline{1-6}
\multirow{6}{*}{3} &  \emph{\textbf{QDDRA$\_$DRL}} &  \emph{\textbf{33.21}} &  \emph{\textbf{34.58}} &  \emph{\textbf{0.14}} & \emph{\textbf{0.79}}       \\
           \cline{2-6}
                    & WMMSE$\_$DRL &33.34 & 34.74 &0.21  &1.05       \\
              \cline{2-6}
                    & QDDRA$\_$BB& 31.96 &33.37  &0.03  &0.14            \\
                \cline{2-6}
                    & WMMSE$\_$BB & 31.94 &33.52  &0.06  &0.30               \\
                  \cline{2-6}
                    & QDDRA$\_$RB & 32.45 &34.95  &0.24  &1.14              \\
                  \cline{2-6}
                    & WMMSE$\_$RB& 32.05 &35.17  &0.37  & 1.74                   \\
\cline{1-6}
\multirow{6}{*}{4} &  \emph{\textbf{QDDRA$\_$DRL}} &  \emph{\textbf{33.11}}  & \emph{\textbf{34.54}}  & \emph{\textbf{0.13}}  & \emph{\textbf{0.84}}      \\
                    \cline{2-6}
                    & WMMSE$\_$DRL & 33.09 &34.63  &0.23  & 1.18           \\
                   \cline{2-6}
                    & QDDRA$\_$BB & 31.70 &33.12  &0.03  &0.11                \\
                  \cline{2-6}
                    & WMMSE$\_$BB & 31.51 & 33.27 &0.11  &0.44               \\
                   \cline{2-6}
                    & QDDRA$\_$RB & 32.36 &34.57  &0.15  &0.79                  \\
                  \cline{2-6}
                    & WMMSE$\_$RB & 31.68 &35.08  &0.40  &1.97              \\
\hline          
\end{tabular}
\label{tab:multicell}
\end{table*}

\begin{table*}[htbp]
\caption{Unfairness Index Achieved by Different Algorithms in Multicell Scenario.}\footnotesize
\centering
\begin{tabular}{|m{2.5cm}<{\centering}|m{1.5cm}<{\centering}|m{1.5cm}<{\centering}|m{1.5cm}<{\centering}|m{1.5cm}<{\centering}|m{1.5cm}<{\centering}|}
\hline 
  \textbf{Algorithm} & \textbf{Cell 1} & \textbf{Cell 2} &  \textbf{Cell 3} &  \textbf{Cell 4} & \textbf{Total}   \\
 \hline
  \emph{\textbf{QDDRA$\_$DRL}} &  \emph{\textbf{0.0642}} &  \emph{\textbf{0.0728}} &  \emph{\textbf{0.0690}} & 	 \emph{\textbf{0.0754}} &  \emph{\textbf{0.0708}}      \\
 \hline
 WMMSE$\_$DRL &	0.0725 & 0.0785 & 0.0811 & 0.0837 & 0.0794\\
 \hline
 QDDRA$\_$BB & 0.0772 & 0.0887 & 0.0857  &  0.0920   & 0.0865  \\
    \hline
  WMMSE$\_$BB & 0.0813 & 0.0901 & 0.0927  & 0.1005  & 0.0917    \\
    \hline
  QDDRA$\_$RB & 0.0854 & 0.0900 &0.0987 & 0.1014&  0.0946      \\
     \hline
  WMMSE$\_$RB & 0.0973 &	 0.0959	& 0.1118 & 0.1124 & 0.1050  \\
\hline          
\end{tabular}
\label{tab:multicellunfairness}
\end{table*}

\section{Conclusion}
In this work, we proposed an optimization framework for video streaming rate adaption over multiuser wireless networks. 
Our framework explicitly considers the effect of physical layer resource allocation on video rate variation, and employs a tailored physical layer algorithm (i.e., QDDRA) to assist optimizing long-term user experience (measured by QoE) on the application layer using a deep reinforcement learning approach. 
Unlike existing short-term cross-layer or long-term but application layer-only QoE maximization methods, the proposed framework offers a simple yet effective way to integrate two tasks across layers, namely, physical layer resource allocation and application layer long-term QoE maximization tasks, respectively.
Extensive simulations show that such cross-layer design is effective and promising.
More importantly, these results suggest that long-term video rate optimization can significantly benefit from judicious cross-layer designs, which may open many doors for future research.

\ifCLASSOPTIONcaptionsoff
  \newpage
\fi

\bibliographystyle{IEEEtran}
\bibliography{JSAC.bib}

\end{document}